\documentclass[12pt,preprint]{aastex}

\begin{document}

\title{High Fill-Out, Extreme Mass Ratio Overcontact Binary Systems. X. The new discovered binary XY Leonis Minoris}

\author{Qian S.-B.\altaffilmark{1,2,3}, Liu L.\altaffilmark{1,2,3}, Zhu L.-Y.\altaffilmark{1,2,3},
He, J.-J.\altaffilmark{1,2,3}, Yang, Y.-G.\altaffilmark{4}, and
Bernasconi, L.\altaffilmark{5}}

\altaffiltext{1}{National Astronomical Observatories/Yunnan
Astronomical Observatory, Chinese Academy of Sciences (CAS), P.O.
Box 110, 650011 Kunming, P.R. China (e-mail: qsb@ynao.ac.cn)}

\altaffiltext{2}{United Laboratory of Optical Astronomy, Chinese
Academy of Sciences (ULOAC), 100012 Beijing, P. R. China}

\altaffiltext{3}{Graduate School of the CAS, 100049 Beijing, P.R.
China}

\altaffiltext{4}{School of Physics and Electric Information, Huaibei
Coal Industry Teachers College, 235000 Huaibei, Anhui Province,
China (e-mail: yygcn@163.com)}

\altaffiltext{5}{Les Engarouines Observatory, F-84570
Malemort-du-Comtat, France (e-mail:
laurent.bernasconi.51@wanadoo.fr)}

\begin{abstract}
The new discovered short-period close binary star, XY LMi, was
monitored photometrically since 2006. It is shown that the light
curves are typical EW-type and show complete eclipses with an
eclipse duration of about 80 minutes. By analyzing the complete B,
V, R, and I light curves with the 2003 version of the W-D code,
photometric solutions were determined. It is discovered that XY LMi
is a high fill-out, extreme mass ratio overcontact binary system
with a mass ratio of $q=0.148$ and a fill-out factor of
$f=74.1\,\%$, suggesting that it is on the late evolutionary stage
of late-type tidal-locked binary stars. As observed in other
overcontact binary stars, evidence for the presence of two dark
spots on both components are given. Based on our 19 epoches of
eclipse times, it is found that the orbital period of the
overcontact binary is decreasing continuously at a rate of
$dP/dt=-1.67\times{10^{-7}}$\,days/year, which may be caused by the
mass transfer from the primary to the secondary or/and angular
momentum loss via magnetic stellar wind. The decrease of the orbital
period may result in the increase of the fill-out, and finally, it
will evolve into a single rapid-rotation star when the fluid surface
reaching the outer critical Roche Lobe.
\end{abstract}


\section{Introduction}

XY LMi (=TYC\,2511-167-1) was recently discovered to be a
short-period eclipsing binary by Bernasconi, L. in the course of
asteroidal light determination (see Bernasconi \& Behrend 2003)
(hereafter B\&B). B\&B illustrated a complete CCD light curve
without filters that shows a typical EW type where the light varies
continuously and the depths of both minima are nearly the same. The
amplitude of the light variability is only 0.38\,magnitude, and the
nearly flat eclipse bottom reveals that it is a total eclipsing
binary. They derived a linear ephemeris,
\begin{equation}
Min. I =JD(Hel.)2452366.884+0.4368897\times{E},
\end{equation}
which indicates XY LMi is a short-period W UMa-type binary star with
a period of 10.49 hours. According to the classification of
Binnendijk (1970), it belongs an A-type system where the transit
minimum is deeper than the occultation one. The present name was
later gave by Kazarovets (2006).

The most popular evolutionary scenario for W UMa-type binary stars
is that they are formed from initially detached systems by angular
momentum loss (AML) via magnetic stellar wind (Vilhu 1982; Guinan \&
Bradstreet 1988; Eggen \& Iben 1989). Both the shrinking of the
Roche lobe via AML and the expanding of the components via the
evolution will result in the massive component star to fill the
critical Roche lobe and the system evolves into an overcontact
binary through a primary-to-secondary mass transfer. At the
overcontact phase, the gradual decrease of mass ratio will make them
finally evolve into rapid rotating single stars (e.g., Van't veer
1997) when the distribution of the angular momentum meets the more
familiar criterion (Hut 1980) that the orbital angular momentum is
less then 3 times the total spin angular momentum, i.e.,
$J_{rot}>1/3J_{orb}$. On the other hand, if the degree of
overcontact is rather high, a dynamical instability will be
encountered and the merge of an overcontact binary star is also
inevitable (Rasio \& Shapiro 1995). Therefore, high fill-out,
extreme mass ratio overcontact binary stars may be the progenitors
of Blue Straggler/FK Com-type stars.

The total and shallow eclipse light minima in the light curves
suggest that XY LMi may be a high fill-out, extreme mass ratio
overcontact binary system. This type of binary stars are at the
late-evolutionary stage of W UMa-type binary systems. Therefore, it
was listed in our series of photometric studies of this kind of
binaries (i.e., Qian \& Yang, 2004 (Paper I); Zhu et al., 2005
(Paper II); Qian et al. 2005a (Paper III); Yang et al., 2005 (Paper
IV); Qian et al. (2005b) (Paper V); Qian et al. (2006) (Paper VI);
Qian et al. (2007) (Paper VII); Qian et al. (2008) (Paper VIII);
Yang et al. (2009) (Paper IX)). In the present paper, photometric
analyses and orbital period studies of the new discovered system, XY
LMi, are presented.

\section{Complete CCD light curves and photometric solutions for XY Leonis Minoris}

\begin{table}
\caption{Coordinates of XY Leonis Minoris, the comparison, and the
check stars. The data come from NOMAD Catalog (Zacharias et al.
2005).}
\begin{center}
\begin{small}
\begin{tabular}{llllllll}\hline\hline
Stars           & NOMAD         & $\alpha_{2000}$        &
$\delta_{2000}$    &B     &V     &R     &(B-V)    \\\hline
XY LMi          & 1221-0223544  & $10^{h}34^{m}12.3^{s}$ & $32^\circ08'51.6"$ &11.750&11.206&10.840&0.544     \\
The comparison  & 1220-0213438  & $10^{h}33^{m}20.2^{s}$ & $32^\circ01'24.6"$ &11.931&11.353&10.960&0.578     \\
The check       & 1221-0223369  & $10^{h}33^{m}30.6^{s}$ & $32^\circ06'49.0"$ &11.790&11.850&11.160&-0.060     \\
\hline\hline
\end{tabular}
\end{small}
\end{center}
\end{table}

Photoelectric observations of XY LMi was made from January 26 to 30,
2008 with the 85-cm reflecting telescope at Xinglong station of the
National Astronomical Observatory. The telescope was equipped with a
primary-focus multicolor CCD photometer where a PI1024 BFT
(Back-illuminated and Frame-Transfer) camera was used (Zhou et al.
2009). It has $1024\times1024$ square pixels, each subtending a
projected angle on the sky of $0."96$ and resulting in a field of
view of $16.'5\times16.'5$. The standard Johnson-Bessel BVRI filters
were used (Zhou et al. 2009) and the integration time for each image
was 9\,s. PHOT (measure magnitudes for a list of stars) of the
aperture photometry package of IRAF \footnote{IRAF (an acronym for
Image Reduction and Analysis Facility) is a collection of software
written at the National Optical Astronomy Observatory (NOAO) geared
towards the reduction of astronomical images in pixel array form.}
was used to reduce the observed images. The coordinates of the
variable star, the comparison star, and the check star are listed in
Table 1. The comparison star we chose was close enough to the
variable that the range of air-mass difference between both stars
was very small ($\approx0.0009$). Therefore, extinction correction
was not made.

\begin{figure}
\begin{center}
\includegraphics[angle=0,scale=1.1]{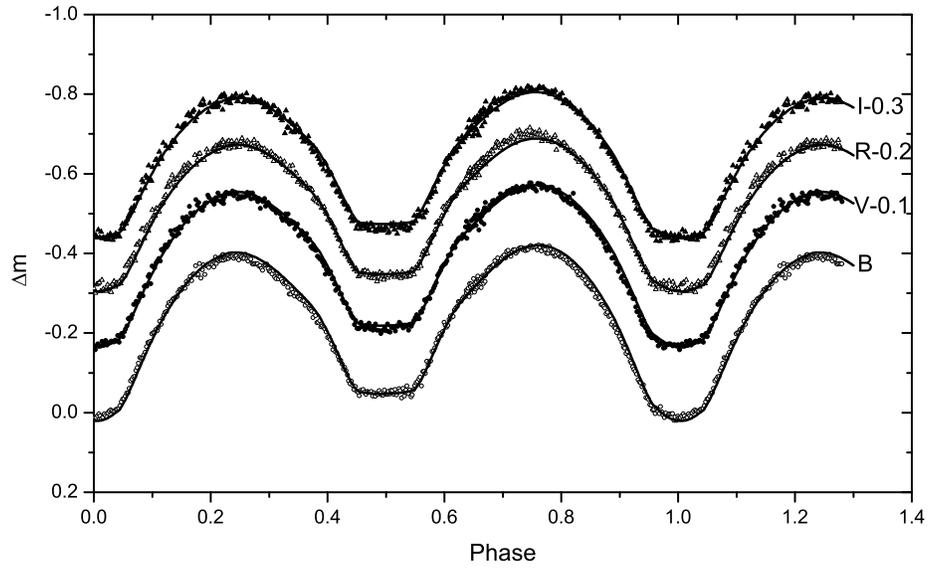}
\caption{CCD photometric light curves in B, V, R, and I bands of XY
Leonis Minoris obtained in January 2008. Also shown in the figure
are the theoretical light curves computed with two dark spots on the
primary and the secondary components, respectively.}
\end{center}
\end{figure}

\begin{table*}
\begin{tiny}
\caption{The original photometric data (HJD-2454400) of XY LMi in B
band.}
\begin{tabular}{llllllllllllll}
\hline
 HJD&$\Delta{m}$&HJD&$\Delta{m}$&HJD&$\Delta{m}$&HJD&$\Delta{m}$&HJD&$\Delta{m}$&HJD&$\Delta{m}$&HJD&$\Delta{m}$\\
 \hline
   92.1825 & $ -.307  $&    92.2543  & $-.393$ &  92.3257  &$ -.003$ &  92.3970 & $ -.276$ &   94.2028 & $ -.386$ &  94.2740 & $ -.122$ & 94.3452 & $ -.186$ \\
   92.1839 & $ -.313  $&    92.2553  & $-.386$ &  92.3267  &$  .002$ &  92.3980 & $ -.273$ &   94.2037 & $ -.381$ &  94.2750 & $ -.110$ & 94.3462 & $ -.189$ \\
   92.1849 & $ -.324  $&    92.2563  & $-.391$ &  92.3277  &$ -.015$ &  92.3990 & $ -.281$ &   94.2047 & $ -.388$ &  94.2760 & $ -.114$ & 94.3472 & $ -.199$ \\
   92.1859 & $ -.323  $&    92.2573  & $-.383$ &  92.3287  &$ -.002$ &  92.4000 & $ -.286$ &   94.2057 & $ -.386$ &  94.2770 & $ -.105$ & 94.3482 & $ -.194$ \\
   92.1869 & $ -.320  $&    92.2583  & $-.376$ &  92.3296  &$  .006$ &  92.4010 & $ -.286$ &   94.2067 & $ -.375$ &  94.2780 & $ -.095$ & 94.3492 & $ -.204$ \\
   92.1879 & $ -.337  $&    92.2593  & $-.374$ &  92.3306  &$  .012$ &  92.4020 & $ -.290$ &   94.2077 & $ -.373$ &  94.2789 & $ -.080$ & 94.3502 & $ -.227$ \\
   92.1889 & $ -.325  $&    92.2603  & $-.376$ &  92.3316  &$  .001$ &  92.4030 & $ -.292$ &   94.2087 & $ -.375$ &  94.2799 & $ -.073$ & 94.3512 & $ -.220$ \\
   92.1899 & $ -.328  $&    92.2613  & $-.368$ &  92.3326  &$  .009$ &  92.4040 & $ -.302$ &   94.2097 & $ -.374$ &  94.2809 & $ -.079$ & 94.3522 & $ -.215$ \\
   92.1909 & $ -.343  $&    92.2623  & $-.365$ &  92.3336  &$  .004$ &  92.4050 & $ -.309$ &   94.2107 & $ -.374$ &  94.2819 & $ -.069$ & 94.3531 & $ -.219$ \\
   92.1919 & $ -.330  $&    92.2632  & $-.369$ &  92.3346  &$  .003$ &  92.4060 & $ -.308$ &   94.2117 & $ -.368$ &  94.2829 & $ -.058$ & 94.3541 & $ -.234$ \\
   92.1929 & $ -.336  $&    92.2642  & $-.362$ &  92.3356  &$  .003$ &  92.4070 & $ -.314$ &   94.2126 & $ -.370$ &  94.2839 & $ -.066$ & 94.3551 & $ -.243$ \\
   92.1939 & $ -.341  $&    92.2652  & $-.354$ &  92.3366  &$  .002$ &  92.4079 & $ -.313$ &   94.2136 & $ -.363$ &  94.2849 & $ -.054$ & 94.3561 & $ -.240$ \\
   92.1949 & $ -.345  $&    92.2662  & $-.346$ &  92.3376  &$  .019$ &  92.4089 & $ -.313$ &   94.2146 & $ -.358$ &  94.2859 & $ -.051$ & 94.3571 & $ -.242$ \\
   92.1958 & $ -.343  $&    92.2672  & $-.349$ &  92.3386  &$  .011$ &  92.4099 & $ -.329$ &   94.2156 & $ -.370$ &  94.2869 & $ -.043$ & 94.3581 & $ -.258$ \\
   92.1968 & $ -.346  $&    92.2682  & $-.357$ &  92.3396  &$  .019$ &  92.4109 & $ -.324$ &   94.2166 & $ -.371$ &  94.2878 & $ -.056$ & 94.3591 & $ -.263$ \\
   92.1978 & $ -.357  $&    92.2692  & $-.339$ &  92.3405  &$  .009$ &  92.4119 & $ -.334$ &   94.2176 & $ -.373$ &  94.2888 & $ -.054$ & 94.3601 & $ -.267$ \\
   92.1988 & $ -.357  $&    92.2702  & $-.344$ &  92.3415  &$  .012$ &  92.4129 & $ -.342$ &   94.2186 & $ -.362$ &  94.2898 & $ -.058$ & 94.3611 & $ -.279$ \\
   92.1998 & $ -.365  $&    92.2712  & $-.333$ &  92.3425  &$  .012$ &  92.4139 & $ -.332$ &   94.2196 & $ -.356$ &  94.2908 & $ -.052$ & 94.3621 & $ -.279$ \\
   92.2008 & $ -.366  $&    92.2722  & $-.323$ &  92.3435  &$  .008$ &  92.4149 & $ -.348$ &   94.2206 & $ -.354$ &  94.2918 & $ -.046$ & 94.3631 & $ -.280$ \\
   92.2018 & $ -.369  $&    92.2731  & $-.326$ &  92.3445  &$  .003$ &  92.4159 & $ -.348$ &   94.2216 & $ -.349$ &  94.2928 & $ -.053$ & 94.3640 & $ -.286$ \\
   92.2028 & $ -.371  $&    92.2741  & $-.317$ &  92.3455  &$  .008$ &  92.4169 & $ -.369$ &   94.2225 & $ -.357$ &  94.2938 & $ -.052$ & 94.3650 & $ -.290$ \\
   92.2038 & $ -.382  $&    92.2751  & $-.318$ &  92.3465  &$  .015$ &  92.4179 & $ -.346$ &   94.2235 & $ -.341$ &  94.2948 & $ -.052$ & 94.3660 & $ -.291$ \\
   92.2048 & $ -.370  $&    92.2761  & $-.310$ &  92.3475  &$  .012$ &  92.4188 & $ -.348$ &   94.2245 & $ -.337$ &  94.2958 & $ -.053$ & 94.3670 & $ -.301$ \\
   92.2058 & $ -.389  $&    92.2771  & $-.310$ &  92.3485  &$  .003$ &  92.4198 & $ -.375$ &   94.2255 & $ -.351$ &  94.2968 & $ -.042$ & 94.3680 & $ -.306$ \\
   92.2067 & $ -.377  $&    92.2781  & $-.309$ &  92.3495  &$  .007$ &  92.4208 & $ -.338$ &   94.2265 & $ -.336$ &  94.2978 & $ -.050$ & 94.3690 & $ -.292$ \\
   92.2077 & $ -.380  $&    92.2791  & $-.290$ &  92.3505  &$  .011$ &  92.4218 & $ -.359$ &   94.2275 & $ -.341$ &  94.2987 & $ -.041$ & 94.3700 & $ -.316$ \\
   92.2087 & $ -.387  $&    92.2801  & $-.300$ &  92.3514  &$  .000$ &  92.4228 & $ -.371$ &   94.2285 & $ -.335$ &  94.2997 & $ -.048$ & 94.3710 & $ -.302$ \\
   92.2097 & $ -.392  $&    92.2811  & $-.286$ &  92.3524  &$ -.004$ &  92.4238 & $ -.347$ &   94.2295 & $ -.329$ &  94.3007 & $ -.046$ & 94.3720 & $ -.314$ \\
   92.2107 & $ -.396  $&    92.2821  & $-.289$ &  92.3534  &$ -.007$ &  92.4248 & $ -.369$ &   94.2305 & $ -.324$ &  94.3017 & $ -.055$ & 94.3730 & $ -.303$ \\
   92.2117 & $ -.389  $&    92.2831  & $-.272$ &  92.3544  &$ -.005$ &  92.4258 & $ -.368$ &   94.2314 & $ -.327$ &  94.3027 & $ -.049$ & 94.3739 & $ -.311$ \\
   92.2127 & $ -.405  $&    92.2841  & $-.271$ &  92.3554  &$ -.002$ &  92.4268 & $ -.365$ &   94.2324 & $ -.327$ &  94.3037 & $ -.044$ & 94.3749 & $ -.315$ \\
   92.2137 & $ -.408  $&    92.2850  & $-.267$ &  92.3564  &$ -.003$ &  92.4278 & $ -.393$ &   94.2334 & $ -.315$ &  94.3047 & $ -.044$ & 94.3759 & $ -.334$ \\
   92.2147 & $ -.398  $&    92.2860  & $-.263$ &  92.3574  &$ -.008$ &  92.4288 & $ -.406$ &   94.2344 & $ -.298$ &  94.3057 & $ -.057$ & 94.3769 & $ -.316$ \\
   92.2157 & $ -.393  $&    92.2870  & $-.249$ &  92.3584  &$ -.008$ &  92.4297 & $ -.413$ &   94.2354 & $ -.298$ &  94.3067 & $ -.046$ & 94.3779 & $ -.314$ \\
   92.2167 & $ -.403  $&    92.2880  & $-.252$ &  92.3594  &$ -.010$ &  92.4307 & $ -.441$ &   94.2364 & $ -.310$ &  94.3077 & $ -.046$ & 94.3789 & $ -.331$ \\
   92.2177 & $ -.409  $&    92.2890  & $-.242$ &  92.3604  &$ -.017$ &  92.4317 & $ -.391$ &   94.2374 & $ -.302$ &  94.3086 & $ -.053$ & 94.3799 & $ -.325$ \\
   92.2186 & $ -.409  $&    92.2900  & $-.244$ &  92.3614  &$ -.027$ &  92.4327 & $ -.405$ &   94.2384 & $ -.293$ &  94.3096 & $ -.047$ & 94.3809 & $ -.330$ \\
   92.2196 & $ -.401  $&    92.2910  & $-.222$ &  92.3624  &$ -.034$ &  92.4337 & $ -.433$ &   94.2394 & $ -.300$ &  94.3106 & $ -.056$ & 96.1046 & $ -.253$ \\
   92.2206 & $ -.405  $&    92.2920  & $-.228$ &  92.3633  &$ -.035$ &  92.4347 & $ -.401$ &   94.2404 & $ -.291$ &  94.3116 & $ -.055$ & 96.1056 & $ -.258$ \\
   92.2216 & $ -.412  $&    92.2930  & $-.215$ &  92.3643  &$ -.041$ &  92.4357 & $ -.405$ &   94.2414 & $ -.290$ &  94.3126 & $ -.046$ & 96.1066 & $ -.266$ \\
   92.2226 & $ -.414  $&    92.2940  & $-.212$ &  92.3653  &$ -.049$ &  92.4367 & $ -.396$ &   94.2423 & $ -.285$ &  94.3136 & $ -.058$ & 96.1076 & $ -.271$ \\
   92.2236 & $ -.415  $&    92.2950  & $-.201$ &  92.3663  &$ -.050$ &  92.4377 & $ -.383$ &   94.2433 & $ -.294$ &  94.3146 & $ -.045$ & 96.1086 & $ -.279$ \\
   92.2246 & $ -.410  $&    92.2960  & $-.209$ &  92.3673  &$ -.061$ &  92.4387 & $ -.388$ &   94.2443 & $ -.282$ &  94.3156 & $ -.052$ & 96.1096 & $ -.280$ \\
   92.2256 & $ -.413  $&    92.2969  & $-.191$ &  92.3683  &$ -.080$ &  92.4397 & $ -.402$ &   94.2453 & $ -.281$ &  94.3166 & $ -.060$ & 96.1106 & $ -.278$ \\
   92.2266 & $ -.407  $&    92.2979  & $-.185$ &  92.3693  &$ -.084$ &  92.4406 & $ -.422$ &   94.2463 & $ -.270$ &  94.3175 & $ -.055$ & 96.1116 & $ -.285$ \\
   92.2276 & $ -.409  $&    92.2989  & $-.176$ &  92.3703  &$ -.090$ &  92.4416 & $ -.324$ &   94.2473 & $ -.277$ &  94.3185 & $ -.037$ & 96.1126 & $ -.287$ \\
   92.2286 & $ -.410  $&    92.2999  & $-.165$ &  92.3713  &$ -.091$ &  94.1770 & $ -.371$ &   94.2483 & $ -.275$ &  94.3195 & $ -.061$ & 96.1135 & $ -.283$ \\
   92.2295 & $ -.417  $&    92.3009  & $-.162$ &  92.3723  &$ -.106$ &  94.1780 & $ -.366$ &   94.2493 & $ -.262$ &  94.3205 & $ -.042$ & 96.1145 & $ -.300$ \\
   92.2305 & $ -.420  $&    92.3019  & $-.149$ &  92.3733  &$ -.113$ &  94.1790 & $ -.376$ &   94.2503 & $ -.259$ &  94.3215 & $ -.057$ & 96.1155 & $ -.298$ \\
   92.2315 & $ -.415  $&    92.3029  & $-.153$ &  92.3742  &$ -.124$ &  94.1800 & $ -.375$ &   94.2513 & $ -.255$ &  94.3225 & $ -.067$ & 96.1165 & $ -.302$ \\
   92.2325 & $ -.416  $&    92.3039  & $-.142$ &  92.3752  &$ -.138$ &  94.1810 & $ -.384$ &   94.2522 & $ -.239$ &  94.3235 & $ -.048$ & 96.1175 & $ -.309$ \\
   92.2335 & $ -.416  $&    92.3049  & $-.129$ &  92.3762  &$ -.137$ &  94.1820 & $ -.382$ &   94.2532 & $ -.252$ &  94.3245 & $ -.057$ & 96.1185 & $ -.304$ \\
   92.2345 & $ -.418  $&    92.3059  & $-.125$ &  92.3772  &$ -.136$ &  94.1830 & $ -.384$ &   94.2542 & $ -.244$ &  94.3255 & $ -.040$ & 96.1195 & $ -.319$ \\
   92.2355 & $ -.409  $&    92.3069  & $-.115$ &  92.3782  &$ -.145$ &  94.1840 & $ -.383$ &   94.2552 & $ -.226$ &  94.3264 & $ -.045$ & 96.1205 & $ -.315$ \\
   92.2365 & $ -.421  $&    92.3078  & $-.109$ &  92.3792  &$ -.152$ &  94.1850 & $ -.397$ &   94.2562 & $ -.227$ &  94.3274 & $ -.058$ & 96.1215 & $ -.334$ \\
   92.2375 & $ -.416  $&    92.3088  & $-.104$ &  92.3802  &$ -.166$ &  94.1859 & $ -.392$ &   94.2572 & $ -.231$ &  94.3284 & $ -.068$ & 96.1225 & $ -.333$ \\
   92.2385 & $ -.406  $&    92.3098  & $-.097$ &  92.3812  &$ -.176$ &  94.1869 & $ -.385$ &   94.2582 & $ -.216$ &  94.3294 & $ -.075$ & 96.1234 & $ -.330$ \\
   92.2395 & $ -.420  $&    92.3108  & $-.083$ &  92.3822  &$ -.176$ &  94.1879 & $ -.388$ &   94.2592 & $ -.218$ &  94.3304 & $ -.078$ & 96.1244 & $ -.329$ \\
   92.2405 & $ -.412  $&    92.3118  & $-.077$ &  92.3832  &$ -.179$ &  94.1889 & $ -.387$ &   94.2602 & $ -.214$ &  94.3314 & $ -.084$ & 96.1254 & $ -.336$ \\
   92.2414 & $ -.410  $&    92.3128  & $-.075$ &  92.3842  &$ -.198$ &  94.1899 & $ -.392$ &   94.2611 & $ -.211$ &  94.3324 & $ -.089$ & 96.1264 & $ -.332$ \\
   92.2424 & $ -.401  $&    92.3138  & $-.065$ &  92.3852  &$ -.201$ &  94.1909 & $ -.392$ &   94.2621 & $ -.203$ &  94.3334 & $ -.067$ & 96.1274 & $ -.345$ \\
   92.2434 & $ -.400  $&    92.3148  & $-.070$ &  92.3861  &$ -.200$ &  94.1919 & $ -.392$ &   94.2631 & $ -.189$ &  94.3344 & $ -.103$ & 96.1284 & $ -.343$ \\
   92.2444 & $ -.418  $&    92.3158  & $-.053$ &  92.3871  &$ -.209$ &  94.1929 & $ -.388$ &   94.2641 & $ -.191$ &  94.3353 & $ -.104$ & 96.1294 & $ -.350$ \\
   92.2454 & $ -.412  $&    92.3168  & $-.038$ &  92.3881  &$ -.206$ &  94.1939 & $ -.392$ &   94.2651 & $ -.184$ &  94.3363 & $ -.099$ & 96.1304 & $ -.354$ \\
   92.2464 & $ -.401  $&    92.3178  & $-.037$ &  92.3891  &$ -.214$ &  94.1948 & $ -.394$ &   94.2661 & $ -.172$ &  94.3373 & $ -.129$ & 96.1314 & $ -.356$ \\
   92.2474 & $ -.402  $&    92.3188  & $-.033$ &  92.3901  &$ -.230$ &  94.1958 & $ -.388$ &   94.2671 & $ -.158$ &  94.3383 & $ -.119$ & 96.1324 & $ -.349$ \\
   92.2484 & $ -.393  $&    92.3197  & $-.020$ &  92.3911  &$ -.222$ &  94.1968 & $ -.384$ &   94.2681 & $ -.166$ &  94.3393 & $ -.128$ & 96.1334 & $ -.357$ \\
   92.2494 & $ -.396  $&    92.3207  & $-.019$ &  92.3921  &$ -.236$ &  94.1978 & $ -.393$ &   94.2691 & $ -.153$ &  94.3403 & $ -.130$ & 96.1343 & $ -.361$ \\
   92.2504 & $ -.402  $&    92.3217  & $-.007$ &  92.3931  &$ -.245$ &  94.1988 & $ -.391$ &   94.2700 & $ -.150$ &  94.3413 & $ -.154$ & 96.1353 & $ -.367$ \\
   92.2514 & $ -.398  $&    92.3227  & $-.008$ &  92.3941  &$ -.252$ &  94.1998 & $ -.387$ &   94.2710 & $ -.141$ &  94.3423 & $ -.155$ & 96.1363 & $ -.374$ \\
   92.2523 & $ -.392  $&    92.3237  & $ .002$ &  92.3951  &$ -.254$ &  94.2008 & $ -.388$ &   94.2720 & $ -.132$ &  94.3433 & $ -.163$ & 96.1373 & $ -.363$ \\
   92.2533 & $ -.391  $&    92.3247  & $-.011$ &  92.3961  &$ -.268$ &  94.2018 & $ -.398$ &   94.2730 & $ -.131$ &  94.3443 & $ -.175$ & 96.1383 & $ -.372$ \\
\hline
\end{tabular}\end{tiny}
\end{table*}

\begin{table*}
\begin{tiny}
\caption{The original photometric data (HJD-2454400) of XY LMi in V
band.}
\begin{tabular}{llllllllllllll}
\hline
 HJD&$\Delta{m}$&HJD&$\Delta{m}$&HJD&$\Delta{m}$&HJD&$\Delta{m}$&HJD&$\Delta{m}$&HJD&$\Delta{m}$&HJD&$\Delta{m}$\\
 \hline
    92.1827 &$  -.381$&  92.2546 &$  -.438$&  92.3260 &$  -.068$&   92.3973 &$  -.329$&  94.2030 &$  -.453$&  94.2743 &$  -.184$&  94.3455 &$  -.245$\\
   92.1842 &$  -.371$&  92.2556 &$  -.434$&  92.3269 &$  -.071$&   92.3983 &$  -.357$&  94.2040 &$  -.445$&  94.2753 &$  -.176$&  94.3465 &$  -.251$\\
   92.1852 &$  -.365$&  92.2566 &$  -.438$&  92.3279 &$  -.074$&   92.3993 &$  -.348$&  94.2050 &$  -.452$&  94.2763 &$  -.177$&  94.3475 &$  -.257$\\
   92.1862 &$  -.382$&  92.2576 &$  -.435$&  92.3289 &$  -.073$&   92.4003 &$  -.345$&  94.2060 &$  -.437$&  94.2772 &$  -.174$&  94.3485 &$  -.265$\\
   92.1872 &$  -.385$&  92.2586 &$  -.428$&  92.3299 &$  -.062$&   92.4013 &$  -.359$&  94.2070 &$  -.452$&  94.2782 &$  -.162$&  94.3495 &$  -.269$\\
   92.1882 &$  -.390$&  92.2595 &$  -.442$&  92.3309 &$  -.064$&   92.4023 &$  -.344$&  94.2080 &$  -.439$&  94.2792 &$  -.160$&  94.3505 &$  -.273$\\
   92.1892 &$  -.397$&  92.2605 &$  -.424$&  92.3319 &$  -.072$&   92.4033 &$  -.362$&  94.2089 &$  -.443$&  94.2802 &$  -.150$&  94.3514 &$  -.275$\\
   92.1902 &$  -.384$&  92.2615 &$  -.427$&  92.3329 &$  -.071$&   92.4042 &$  -.350$&  94.2099 &$  -.438$&  94.2812 &$  -.143$&  94.3524 &$  -.281$\\
   92.1912 &$  -.396$&  92.2625 &$  -.451$&  92.3339 &$  -.073$&   92.4052 &$  -.367$&  94.2109 &$  -.430$&  94.2822 &$  -.138$&  94.3534 &$  -.295$\\
   92.1922 &$  -.395$&  92.2635 &$  -.423$&  92.3349 &$  -.073$&   92.4062 &$  -.364$&  94.2119 &$  -.438$&  94.2832 &$  -.121$&  94.3544 &$  -.296$\\
   92.1931 &$  -.398$&  92.2645 &$  -.417$&  92.3359 &$  -.071$&   92.4072 &$  -.390$&  94.2129 &$  -.413$&  94.2842 &$  -.132$&  94.3554 &$  -.297$\\
   92.1941 &$  -.394$&  92.2655 &$  -.415$&  92.3369 &$  -.068$&   92.4082 &$  -.382$&  94.2139 &$  -.433$&  94.2851 &$  -.112$&  94.3564 &$  -.313$\\
   92.1951 &$  -.407$&  92.2665 &$  -.414$&  92.3378 &$  -.071$&   92.4092 &$  -.382$&  94.2149 &$  -.430$&  94.2861 &$  -.111$&  94.3574 &$  -.319$\\
   92.1961 &$  -.415$&  92.2675 &$  -.402$&  92.3388 &$  -.070$&   92.4102 &$  -.392$&  94.2159 &$  -.438$&  94.2871 &$  -.113$&  94.3584 &$  -.315$\\
   92.1971 &$  -.414$&  92.2685 &$  -.404$&  92.3398 &$  -.067$&   92.4112 &$  -.380$&  94.2169 &$  -.419$&  94.2881 &$  -.127$&  94.3594 &$  -.329$\\
   92.1981 &$  -.410$&  92.2695 &$  -.401$&  92.3408 &$  -.071$&   92.4122 &$  -.418$&  94.2179 &$  -.425$&  94.2891 &$  -.117$&  94.3603 &$  -.333$\\
   92.1991 &$  -.410$&  92.2704 &$  -.395$&  92.3418 &$  -.058$&   92.4132 &$  -.401$&  94.2188 &$  -.428$&  94.2901 &$  -.116$&  94.3613 &$  -.338$\\
   92.2001 &$  -.421$&  92.2714 &$  -.396$&  92.3428 &$  -.068$&   92.4142 &$  -.397$&  94.2198 &$  -.422$&  94.2911 &$  -.109$&  94.3623 &$  -.340$\\
   92.2011 &$  -.423$&  92.2724 &$  -.391$&  92.3438 &$  -.073$&   92.4151 &$  -.396$&  94.2208 &$  -.416$&  94.2921 &$  -.119$&  94.3633 &$  -.337$\\
   92.2021 &$  -.432$&  92.2734 &$  -.380$&  92.3448 &$  -.067$&   92.4161 &$  -.403$&  94.2218 &$  -.412$&  94.2931 &$  -.115$&  94.3643 &$  -.348$\\
   92.2031 &$  -.433$&  92.2744 &$  -.386$&  92.3458 &$  -.068$&   92.4171 &$  -.435$&  94.2228 &$  -.410$&  94.2941 &$  -.125$&  94.3653 &$  -.343$\\
   92.2040 &$  -.427$&  92.2754 &$  -.378$&  92.3468 &$  -.079$&   92.4181 &$  -.422$&  94.2238 &$  -.410$&  94.2950 &$  -.110$&  94.3663 &$  -.367$\\
   92.2050 &$  -.431$&  92.2764 &$  -.368$&  92.3478 &$  -.069$&   92.4191 &$  -.428$&  94.2248 &$  -.402$&  94.2960 &$  -.112$&  94.3673 &$  -.338$\\
   92.2060 &$  -.437$&  92.2774 &$  -.367$&  92.3487 &$  -.075$&   92.4201 &$  -.414$&  94.2258 &$  -.404$&  94.2970 &$  -.107$&  94.3683 &$  -.359$\\
   92.2070 &$  -.439$&  92.2784 &$  -.361$&  92.3497 &$  -.081$&   92.4211 &$  -.410$&  94.2268 &$  -.403$&  94.2980 &$  -.116$&  94.3692 &$  -.359$\\
   92.2080 &$  -.450$&  92.2794 &$  -.358$&  92.3507 &$  -.070$&   92.4221 &$  -.408$&  94.2277 &$  -.399$&  94.2990 &$  -.123$&  94.3702 &$  -.363$\\
   92.2090 &$  -.446$&  92.2804 &$  -.347$&  92.3517 &$  -.074$&   92.4231 &$  -.419$&  94.2287 &$  -.389$&  94.3000 &$  -.107$&  94.3712 &$  -.366$\\
   92.2100 &$  -.447$&  92.2813 &$  -.358$&  92.3527 &$  -.077$&   92.4241 &$  -.426$&  94.2297 &$  -.383$&  94.3010 &$  -.088$&  94.3722 &$  -.369$\\
   92.2110 &$  -.446$&  92.2823 &$  -.348$&  92.3537 &$  -.080$&   92.4251 &$  -.427$&  94.2307 &$  -.387$&  94.3020 &$  -.106$&  94.3732 &$  -.370$\\
   92.2120 &$  -.450$&  92.2833 &$  -.337$&  92.3547 &$  -.079$&   92.4260 &$  -.446$&  94.2317 &$  -.399$&  94.3030 &$  -.099$&  94.3742 &$  -.370$\\
   92.2130 &$  -.456$&  92.2843 &$  -.338$&  92.3557 &$  -.078$&   92.4270 &$  -.441$&  94.2327 &$  -.381$&  94.3039 &$  -.105$&  94.3752 &$  -.409$\\
   92.2139 &$  -.455$&  92.2853 &$  -.327$&  92.3567 &$  -.082$&   92.4280 &$  -.487$&  94.2337 &$  -.399$&  94.3049 &$  -.108$&  94.3762 &$  -.359$\\
   92.2149 &$  -.459$&  92.2863 &$  -.315$&  92.3577 &$  -.080$&   92.4290 &$  -.431$&  94.2347 &$  -.374$&  94.3059 &$  -.112$&  94.3772 &$  -.387$\\
   92.2159 &$  -.451$&  92.2873 &$  -.307$&  92.3586 &$  -.094$&   92.4300 &$  -.467$&  94.2357 &$  -.382$&  94.3069 &$  -.106$&  94.3782 &$  -.379$\\
   92.2169 &$  -.463$&  92.2883 &$  -.317$&  92.3596 &$  -.077$&   92.4310 &$  -.410$&  94.2367 &$  -.377$&  94.3079 &$  -.123$&  94.3792 &$  -.402$\\
   92.2179 &$  -.453$&  92.2893 &$  -.313$&  92.3606 &$  -.094$&   92.4320 &$  -.440$&  94.2377 &$  -.378$&  94.3089 &$  -.104$&  94.3801 &$  -.374$\\
   92.2189 &$  -.463$&  92.2903 &$  -.307$&  92.3616 &$  -.099$&   92.4330 &$  -.465$&  94.2386 &$  -.362$&  94.3099 &$  -.102$&  94.3811 &$  -.428$\\
   92.2199 &$  -.468$&  92.2913 &$  -.288$&  92.3626 &$  -.091$&   92.4340 &$  -.391$&  94.2396 &$  -.368$&  94.3109 &$  -.114$&  96.1049 &$  -.307$\\
   92.2209 &$  -.455$&  92.2922 &$  -.290$&  92.3636 &$  -.103$&   92.4350 &$  -.471$&  94.2406 &$  -.374$&  94.3119 &$  -.115$&  96.1059 &$  -.316$\\
   92.2219 &$  -.463$&  92.2932 &$  -.289$&  92.3646 &$  -.110$&   92.4360 &$  -.466$&  94.2416 &$  -.344$&  94.3129 &$  -.113$&  96.1069 &$  -.321$\\
   92.2229 &$  -.458$&  92.2942 &$  -.276$&  92.3656 &$  -.118$&   92.4369 &$  -.427$&  94.2426 &$  -.348$&  94.3138 &$  -.105$&  96.1079 &$  -.317$\\
   92.2239 &$  -.467$&  92.2952 &$  -.273$&  92.3666 &$  -.136$&   92.4379 &$  -.425$&  94.2436 &$  -.357$&  94.3148 &$  -.118$&  96.1088 &$  -.339$\\
   92.2249 &$  -.469$&  92.2962 &$  -.255$&  92.3676 &$  -.138$&   92.4389 &$  -.450$&  94.2446 &$  -.352$&  94.3158 &$  -.117$&  96.1098 &$  -.334$\\
   92.2258 &$  -.459$&  92.2972 &$  -.264$&  92.3686 &$  -.137$&   92.4399 &$  -.517$&  94.2456 &$  -.345$&  94.3168 &$  -.127$&  96.1108 &$  -.347$\\
   92.2268 &$  -.467$&  92.2982 &$  -.238$&  92.3695 &$  -.146$&   92.4409 &$  -.370$&  94.2466 &$  -.335$&  94.3178 &$  -.123$&  96.1118 &$  -.352$\\
   92.2278 &$  -.468$&  92.2992 &$  -.243$&  92.3705 &$  -.172$&   92.4419 &$  -.467$&  94.2475 &$  -.350$&  94.3188 &$  -.118$&  96.1128 &$  -.354$\\
   92.2288 &$  -.467$&  92.3002 &$  -.229$&  92.3715 &$  -.173$&   94.1773 &$  -.435$&  94.2485 &$  -.331$&  94.3198 &$  -.113$&  96.1138 &$  -.339$\\
   92.2298 &$  -.477$&  92.3012 &$  -.228$&  92.3725 &$  -.180$&   94.1783 &$  -.424$&  94.2495 &$  -.336$&  94.3208 &$  -.120$&  96.1148 &$  -.360$\\
   92.2308 &$  -.477$&  92.3022 &$  -.223$&  92.3735 &$  -.178$&   94.1793 &$  -.433$&  94.2505 &$  -.317$&  94.3218 &$  -.117$&  96.1158 &$  -.357$\\
   92.2318 &$  -.465$&  92.3031 &$  -.218$&  92.3745 &$  -.192$&   94.1803 &$  -.434$&  94.2515 &$  -.313$&  94.3227 &$  -.118$&  96.1168 &$  -.372$\\
   92.2328 &$  -.464$&  92.3041 &$  -.200$&  92.3755 &$  -.199$&   94.1813 &$  -.452$&  94.2525 &$  -.321$&  94.3237 &$  -.107$&  96.1178 &$  -.366$\\
   92.2338 &$  -.462$&  92.3051 &$  -.206$&  92.3765 &$  -.195$&   94.1822 &$  -.445$&  94.2535 &$  -.309$&  94.3247 &$  -.122$&  96.1188 &$  -.364$\\
   92.2348 &$  -.457$&  92.3061 &$  -.199$&  92.3775 &$  -.194$&   94.1832 &$  -.431$&  94.2545 &$  -.308$&  94.3257 &$  -.117$&  96.1197 &$  -.375$\\
   92.2358 &$  -.476$&  92.3071 &$  -.187$&  92.3785 &$  -.217$&   94.1842 &$  -.444$&  94.2555 &$  -.293$&  94.3267 &$  -.124$&  96.1207 &$  -.376$\\
   92.2367 &$  -.473$&  92.3081 &$  -.187$&  92.3795 &$  -.220$&   94.1852 &$  -.435$&  94.2565 &$  -.290$&  94.3277 &$  -.131$&  96.1217 &$  -.381$\\
   92.2377 &$  -.475$&  92.3091 &$  -.166$&  92.3805 &$  -.218$&   94.1862 &$  -.442$&  94.2574 &$  -.286$&  94.3287 &$  -.139$&  96.1227 &$  -.375$\\
   92.2387 &$  -.464$&  92.3101 &$  -.154$&  92.3814 &$  -.242$&   94.1872 &$  -.453$&  94.2584 &$  -.286$&  94.3297 &$  -.146$&  96.1237 &$  -.394$\\
   92.2397 &$  -.472$&  92.3111 &$  -.154$&  92.3824 &$  -.247$&   94.1882 &$  -.452$&  94.2594 &$  -.278$&  94.3307 &$  -.136$&  96.1247 &$  -.393$\\
   92.2407 &$  -.469$&  92.3121 &$  -.143$&  92.3834 &$  -.224$&   94.1892 &$  -.450$&  94.2604 &$  -.269$&  94.3317 &$  -.136$&  96.1257 &$  -.391$\\
   92.2417 &$  -.465$&  92.3131 &$  -.132$&  92.3844 &$  -.268$&   94.1902 &$  -.451$&  94.2614 &$  -.248$&  94.3326 &$  -.154$&  96.1267 &$  -.398$\\
   92.2427 &$  -.460$&  92.3141 &$  -.122$&  92.3854 &$  -.266$&   94.1912 &$  -.450$&  94.2624 &$  -.260$&  94.3336 &$  -.152$&  96.1277 &$  -.401$\\
   92.2437 &$  -.463$&  92.3150 &$  -.131$&  92.3864 &$  -.264$&   94.1921 &$  -.457$&  94.2634 &$  -.256$&  94.3346 &$  -.162$&  96.1287 &$  -.405$\\
   92.2447 &$  -.465$&  92.3160 &$  -.116$&  92.3874 &$  -.276$&   94.1931 &$  -.445$&  94.2644 &$  -.249$&  94.3356 &$  -.183$&  96.1297 &$  -.410$\\
   92.2457 &$  -.467$&  92.3170 &$  -.112$&  92.3884 &$  -.282$&   94.1941 &$  -.453$&  94.2654 &$  -.254$&  94.3366 &$  -.180$&  96.1307 &$  -.408$\\
   92.2467 &$  -.466$&  92.3180 &$  -.103$&  92.3894 &$  -.287$&   94.1951 &$  -.451$&  94.2664 &$  -.236$&  94.3376 &$  -.186$&  96.1316 &$  -.419$\\
   92.2477 &$  -.457$&  92.3190 &$  -.094$&  92.3904 &$  -.281$&   94.1961 &$  -.437$&  94.2673 &$  -.238$&  94.3386 &$  -.189$&  96.1326 &$  -.418$\\
   92.2486 &$  -.457$&  92.3200 &$  -.099$&  92.3914 &$  -.310$&   94.1971 &$  -.443$&  94.2683 &$  -.223$&  94.3396 &$  -.187$&  96.1336 &$  -.422$\\
   92.2496 &$  -.454$&  92.3210 &$  -.088$&  92.3924 &$  -.307$&   94.1981 &$  -.439$&  94.2693 &$  -.207$&  94.3406 &$  -.202$&  96.1346 &$  -.417$\\
   92.2506 &$  -.457$&  92.3220 &$  -.088$&  92.3933 &$  -.313$&   94.1991 &$  -.443$&  94.2703 &$  -.212$&  94.3416 &$  -.205$&  96.1356 &$  -.422$\\
   92.2516 &$  -.460$&  92.3230 &$  -.088$&  92.3943 &$  -.321$&   94.2000 &$  -.442$&  94.2713 &$  -.215$&  94.3425 &$  -.213$&  96.1366 &$  -.432$\\
   92.2526 &$  -.450$&  92.3240 &$  -.086$&  92.3953 &$  -.316$&   94.2010 &$  -.442$&  94.2723 &$  -.198$&  94.3435 &$  -.219$&  96.1376 &$  -.425$\\
   92.2536 &$  -.443$&  92.3250 &$  -.081$&  92.3963 &$  -.322$&   94.2020 &$  -.445$&  94.2733 &$  -.198$&  94.3445 &$  -.227$&  96.1386 &$  -.422$\\
\hline
\end{tabular}\end{tiny}
\end{table*}

\begin{table*}
\begin{tiny}
\caption{The original photometric data (HJD-2454400) of XY LMi in R
band.}
\begin{tabular}{llllllllllllll}
\hline
 HJD&$\Delta{m}$&HJD&$\Delta{m}$&HJD&$\Delta{m}$&HJD&$\Delta{m}$&HJD&$\Delta{m}$&HJD&$\Delta{m}$&HJD&$\Delta{m}$\\
 \hline
   92.1830  &$ -.395 $&     92.2548  &$ -.477 $&  92.3262  &$ -.116 $&    92.3975  &$ -.381 $&   94.2072  &$ -.469 $&  94.2784  &$ -.191 $&  94.3497  &$ -.317 $\\
   92.1844  &$ -.407 $&     92.2558  &$ -.478 $&  92.3272  &$ -.125 $&    92.3985  &$ -.373 $&   94.2082  &$ -.485 $&  94.2794  &$ -.184 $&  94.3507  &$ -.313 $\\
   92.1854  &$ -.414 $&     92.2568  &$ -.464 $&  92.3282  &$ -.122 $&    92.3995  &$ -.396 $&   94.2092  &$ -.470 $&  94.2804  &$ -.180 $&  94.3517  &$ -.309 $\\
   92.1864  &$ -.405 $&     92.2578  &$ -.468 $&  92.3291  &$ -.111 $&    92.4005  &$ -.385 $&   94.2102  &$ -.463 $&  94.2814  &$ -.175 $&  94.3527  &$ -.325 $\\
   92.1874  &$ -.404 $&     92.2588  &$ -.458 $&  92.3301  &$ -.130 $&    92.4015  &$ -.370 $&   94.2112  &$ -.475 $&  94.2824  &$ -.167 $&  94.3536  &$ -.324 $\\
   92.1884  &$ -.417 $&     92.2598  &$ -.457 $&  92.3311  &$ -.120 $&    92.4025  &$ -.387 $&   94.2121  &$ -.468 $&  94.2834  &$ -.159 $&  94.3546  &$ -.333 $\\
   92.1894  &$ -.404 $&     92.2608  &$ -.458 $&  92.3321  &$ -.123 $&    92.4035  &$ -.387 $&   94.2131  &$ -.464 $&  94.2844  &$ -.162 $&  94.3556  &$ -.337 $\\
   92.1904  &$ -.424 $&     92.2617  &$ -.452 $&  92.3331  &$ -.107 $&    92.4045  &$ -.398 $&   94.2141  &$ -.463 $&  94.2854  &$ -.154 $&  94.3566  &$ -.340 $\\
   92.1914  &$ -.427 $&     92.2627  &$ -.454 $&  92.3341  &$ -.131 $&    92.4055  &$ -.396 $&   94.2151  &$ -.454 $&  94.2864  &$ -.142 $&  94.3576  &$ -.342 $\\
   92.1924  &$ -.417 $&     92.2637  &$ -.449 $&  92.3351  &$ -.102 $&    92.4065  &$ -.406 $&   94.2161  &$ -.463 $&  94.2873  &$ -.146 $&  94.3586  &$ -.344 $\\
   92.1934  &$ -.436 $&     92.2647  &$ -.440 $&  92.3361  &$ -.112 $&    92.4074  &$ -.410 $&   94.2171  &$ -.447 $&  94.2883  &$ -.149 $&  94.3596  &$ -.350 $\\
   92.1944  &$ -.427 $&     92.2657  &$ -.436 $&  92.3371  &$ -.128 $&    92.4084  &$ -.421 $&   94.2181  &$ -.456 $&  94.2893  &$ -.155 $&  94.3606  &$ -.360 $\\
   92.1953  &$ -.439 $&     92.2667  &$ -.430 $&  92.3381  &$ -.118 $&    92.4094  &$ -.413 $&   94.2191  &$ -.443 $&  94.2903  &$ -.155 $&  94.3616  &$ -.359 $\\
   92.1963  &$ -.438 $&     92.2677  &$ -.439 $&  92.3391  &$ -.113 $&    92.4104  &$ -.407 $&   94.2201  &$ -.451 $&  94.2913  &$ -.157 $&  94.3626  &$ -.371 $\\
   92.1973  &$ -.440 $&     92.2687  &$ -.434 $&  92.3400  &$ -.107 $&    92.4114  &$ -.426 $&   94.2211  &$ -.455 $&  94.2923  &$ -.146 $&  94.3635  &$ -.369 $\\
   92.1983  &$ -.450 $&     92.2697  &$ -.432 $&  92.3410  &$ -.112 $&    92.4124  &$ -.422 $&   94.2220  &$ -.449 $&  94.2933  &$ -.154 $&  94.3645  &$ -.371 $\\
   92.1993  &$ -.438 $&     92.2707  &$ -.418 $&  92.3420  &$ -.101 $&    92.4134  &$ -.419 $&   94.2230  &$ -.432 $&  94.2943  &$ -.140 $&  94.3655  &$ -.382 $\\
   92.2003  &$ -.450 $&     92.2717  &$ -.422 $&  92.3430  &$ -.122 $&    92.4144  &$ -.306 $&   94.2240  &$ -.444 $&  94.2953  &$ -.146 $&  94.3665  &$ -.364 $\\
   92.2013  &$ -.459 $&     92.2727  &$ -.407 $&  92.3440  &$ -.124 $&    92.4154  &$ -.466 $&   94.2250  &$ -.441 $&  94.2963  &$ -.138 $&  94.3675  &$ -.378 $\\
   92.2023  &$ -.456 $&     92.2736  &$ -.413 $&  92.3450  &$ -.127 $&    92.4164  &$ -.447 $&   94.2260  &$ -.437 $&  94.2973  &$ -.148 $&  94.3685  &$ -.394 $\\
   92.2033  &$ -.465 $&     92.2746  &$ -.417 $&  92.3460  &$ -.105 $&    92.4174  &$ -.423 $&   94.2270  &$ -.432 $&  94.2982  &$ -.138 $&  94.3695  &$ -.403 $\\
   92.2043  &$ -.468 $&     92.2756  &$ -.404 $&  92.3470  &$ -.133 $&    92.4183  &$ -.434 $&   94.2280  &$ -.436 $&  94.2992  &$ -.147 $&  94.3705  &$ -.392 $\\
   92.2053  &$ -.480 $&     92.2766  &$ -.405 $&  92.3480  &$ -.111 $&    92.4193  &$ -.450 $&   94.2290  &$ -.426 $&  94.3002  &$ -.130 $&  94.3715  &$ -.404 $\\
   92.2063  &$ -.478 $&     92.2776  &$ -.393 $&  92.3490  &$ -.107 $&    92.4203  &$ -.451 $&   94.2300  &$ -.422 $&  94.3012  &$ -.141 $&  94.3725  &$ -.409 $\\
   92.2072  &$ -.472 $&     92.2786  &$ -.393 $&  92.3500  &$ -.106 $&    92.4213  &$ -.434 $&   94.2309  &$ -.431 $&  94.3022  &$ -.153 $&  94.3734  &$ -.389 $\\
   92.2082  &$ -.465 $&     92.2796  &$ -.382 $&  92.3510  &$ -.121 $&    92.4223  &$ -.471 $&   94.2319  &$ -.409 $&  94.3032  &$ -.136 $&  94.3744  &$ -.420 $\\
   92.2092  &$ -.475 $&     92.2806  &$ -.395 $&  92.3519  &$ -.111 $&    92.4233  &$ -.435 $&   94.2329  &$ -.453 $&  94.3042  &$ -.149 $&  94.3754  &$ -.413 $\\
   92.2102  &$ -.482 $&     92.2816  &$ -.381 $&  92.3529  &$ -.117 $&    92.4243  &$ -.473 $&   94.2339  &$ -.404 $&  94.3052  &$ -.134 $&  94.3764  &$ -.396 $\\
   92.2112  &$ -.483 $&     92.2826  &$ -.374 $&  92.3539  &$ -.103 $&    92.4253  &$ -.461 $&   94.2349  &$ -.419 $&  94.3062  &$ -.142 $&  94.3774  &$ -.422 $\\
   92.2122  &$ -.480 $&     92.2836  &$ -.365 $&  92.3549  &$ -.123 $&    92.4263  &$ -.454 $&   94.2359  &$ -.403 $&  94.3072  &$ -.143 $&  94.3784  &$ -.375 $\\
   92.2132  &$ -.482 $&     92.2845  &$ -.361 $&  92.3559  &$ -.130 $&    92.4273  &$ -.434 $&   94.2369  &$ -.421 $&  94.3081  &$ -.139 $&  94.3794  &$ -.439 $\\
   92.2142  &$ -.481 $&     92.2855  &$ -.368 $&  92.3569  &$ -.116 $&    92.4283  &$ -.512 $&   94.2379  &$ -.395 $&  94.3091  &$ -.145 $&  94.3804  &$ -.426 $\\
   92.2152  &$ -.478 $&     92.2865  &$ -.364 $&  92.3579  &$ -.126 $&    92.4292  &$ -.434 $&   94.2389  &$ -.403 $&  94.3101  &$ -.152 $&  94.3814  &$ -.424 $\\
   92.2162  &$ -.495 $&     92.2875  &$ -.351 $&  92.3589  &$ -.123 $&    92.4302  &$ -.497 $&   94.2399  &$ -.401 $&  94.3111  &$ -.166 $&  96.1051  &$ -.346 $\\
   92.2172  &$ -.491 $&     92.2885  &$ -.340 $&  92.3599  &$ -.129 $&    92.4312  &$ -.465 $&   94.2409  &$ -.380 $&  94.3121  &$ -.147 $&  96.1061  &$ -.347 $\\
   92.2181  &$ -.488 $&     92.2895  &$ -.347 $&  92.3609  &$ -.125 $&    92.4322  &$ -.471 $&   94.2418  &$ -.385 $&  94.3131  &$ -.141 $&  96.1071  &$ -.347 $\\
   92.2191  &$ -.478 $&     92.2905  &$ -.337 $&  92.3619  &$ -.138 $&    92.4332  &$ -.492 $&   94.2428  &$ -.386 $&  94.3141  &$ -.146 $&  96.1081  &$ -.342 $\\
   92.2201  &$ -.506 $&     92.2915  &$ -.330 $&  92.3628  &$ -.144 $&    92.4352  &$ -.500 $&   94.2438  &$ -.373 $&  94.3151  &$ -.152 $&  96.1091  &$ -.369 $\\
   92.2211  &$ -.491 $&     92.2925  &$ -.320 $&  92.3638  &$ -.144 $&    92.4362  &$ -.476 $&   94.2448  &$ -.376 $&  94.3161  &$ -.147 $&  96.1101  &$ -.358 $\\
   92.2221  &$ -.509 $&     92.2935  &$ -.319 $&  92.3648  &$ -.168 $&    92.4372  &$ -.475 $&   94.2458  &$ -.370 $&  94.3170  &$ -.141 $&  96.1111  &$ -.367 $\\
   92.2231  &$ -.501 $&     92.2945  &$ -.297 $&  92.3658  &$ -.172 $&    92.4392  &$ -.497 $&   94.2468  &$ -.376 $&  94.3180  &$ -.138 $&  96.1120  &$ -.378 $\\
   92.2241  &$ -.498 $&     92.2955  &$ -.300 $&  92.3668  &$ -.178 $&    92.4421  &$ -.463 $&   94.2478  &$ -.365 $&  94.3190  &$ -.153 $&  96.1130  &$ -.373 $\\
   92.2251  &$ -.491 $&     92.2964  &$ -.304 $&  92.3678  &$ -.185 $&    94.1775  &$ -.462 $&   94.2488  &$ -.362 $&  94.3200  &$ -.144 $&  96.1140  &$ -.384 $\\
   92.2261  &$ -.500 $&     92.2974  &$ -.295 $&  92.3688  &$ -.184 $&    94.1785  &$ -.462 $&   94.2498  &$ -.363 $&  94.3210  &$ -.146 $&  96.1150  &$ -.387 $\\
   92.2271  &$ -.506 $&     92.2984  &$ -.269 $&  92.3698  &$ -.194 $&    94.1795  &$ -.469 $&   94.2508  &$ -.362 $&  94.3220  &$ -.141 $&  96.1160  &$ -.392 $\\
   92.2281  &$ -.500 $&     92.2994  &$ -.289 $&  92.3708  &$ -.200 $&    94.1805  &$ -.455 $&   94.2517  &$ -.360 $&  94.3230  &$ -.150 $&  96.1170  &$ -.379 $\\
   92.2291  &$ -.496 $&     92.3004  &$ -.287 $&  92.3718  &$ -.182 $&    94.1815  &$ -.472 $&   94.2527  &$ -.338 $&  94.3240  &$ -.143 $&  96.1180  &$ -.398 $\\
   92.2300  &$ -.516 $&     92.3014  &$ -.264 $&  92.3728  &$ -.222 $&    94.1825  &$ -.475 $&   94.2537  &$ -.335 $&  94.3250  &$ -.144 $&  96.1190  &$ -.404 $\\
   92.2310  &$ -.496 $&     92.3024  &$ -.257 $&  92.3738  &$ -.222 $&    94.1835  &$ -.472 $&   94.2547  &$ -.331 $&  94.3259  &$ -.159 $&  96.1200  &$ -.416 $\\
   92.2320  &$ -.505 $&     92.3034  &$ -.254 $&  92.3747  &$ -.229 $&    94.1845  &$ -.484 $&   94.2557  &$ -.338 $&  94.3269  &$ -.158 $&  96.1210  &$ -.404 $\\
   92.2330  &$ -.502 $&     92.3044  &$ -.236 $&  92.3757  &$ -.226 $&    94.1855  &$ -.480 $&   94.2567  &$ -.327 $&  94.3279  &$ -.147 $&  96.1220  &$ -.397 $\\
   92.2340  &$ -.503 $&     92.3054  &$ -.232 $&  92.3767  &$ -.261 $&    94.1864  &$ -.470 $&   94.2577  &$ -.324 $&  94.3289  &$ -.164 $&  96.1229  &$ -.407 $\\
   92.2350  &$ -.499 $&     92.3064  &$ -.224 $&  92.3777  &$ -.261 $&    94.1874  &$ -.466 $&   94.2587  &$ -.315 $&  94.3299  &$ -.162 $&  96.1239  &$ -.421 $\\
   92.2360  &$ -.498 $&     92.3074  &$ -.217 $&  92.3787  &$ -.238 $&    94.1884  &$ -.481 $&   94.2597  &$ -.314 $&  94.3309  &$ -.163 $&  96.1249  &$ -.422 $\\
   92.2370  &$ -.509 $&     92.3083  &$ -.216 $&  92.3797  &$ -.256 $&    94.1894  &$ -.483 $&   94.2606  &$ -.317 $&  94.3319  &$ -.178 $&  96.1259  &$ -.431 $\\
   92.2380  &$ -.498 $&     92.3093  &$ -.206 $&  92.3807  &$ -.270 $&    94.1904  &$ -.473 $&   94.2616  &$ -.304 $&  94.3329  &$ -.168 $&  96.1269  &$ -.422 $\\
   92.2390  &$ -.482 $&     92.3103  &$ -.197 $&  92.3817  &$ -.257 $&    94.1914  &$ -.482 $&   94.2626  &$ -.299 $&  94.3339  &$ -.196 $&  96.1279  &$ -.429 $\\
   92.2400  &$ -.493 $&     92.3113  &$ -.199 $&  92.3827  &$ -.274 $&    94.1924  &$ -.472 $&   94.2636  &$ -.285 $&  94.3348  &$ -.205 $&  96.1289  &$ -.435 $\\
   92.2409  &$ -.488 $&     92.3123  &$ -.189 $&  92.3837  &$ -.284 $&    94.1934  &$ -.466 $&   94.2646  &$ -.284 $&  94.3358  &$ -.198 $&  96.1299  &$ -.441 $\\
   92.2419  &$ -.503 $&     92.3133  &$ -.180 $&  92.3847  &$ -.309 $&    94.1944  &$ -.474 $&   94.2656  &$ -.281 $&  94.3368  &$ -.210 $&  96.1309  &$ -.435 $\\
   92.2429  &$ -.488 $&     92.3143  &$ -.172 $&  92.3856  &$ -.295 $&    94.1953  &$ -.483 $&   94.2666  &$ -.275 $&  94.3378  &$ -.200 $&  96.1319  &$ -.442 $\\
   92.2439  &$ -.482 $&     92.3153  &$ -.158 $&  92.3866  &$ -.294 $&    94.1963  &$ -.469 $&   94.2676  &$ -.266 $&  94.3388  &$ -.231 $&  96.1329  &$ -.452 $\\
   92.2449  &$ -.489 $&     92.3163  &$ -.165 $&  92.3876  &$ -.313 $&    94.1973  &$ -.467 $&   94.2686  &$ -.263 $&  94.3398  &$ -.227 $&  96.1338  &$ -.449 $\\
   92.2459  &$ -.482 $&     92.3173  &$ -.152 $&  92.3886  &$ -.312 $&    94.1983  &$ -.482 $&   94.2695  &$ -.252 $&  94.3408  &$ -.241 $&  96.1348  &$ -.462 $\\
   92.2469  &$ -.479 $&     92.3182  &$ -.153 $&  92.3896  &$ -.305 $&    94.1993  &$ -.489 $&   94.2705  &$ -.244 $&  94.3418  &$ -.244 $&  96.1358  &$ -.457 $\\
   92.2479  &$ -.492 $&     92.3192  &$ -.140 $&  92.3906  &$ -.320 $&    94.2003  &$ -.477 $&   94.2715  &$ -.236 $&  94.3428  &$ -.248 $&  96.1368  &$ -.458 $\\
   92.2489  &$ -.496 $&     92.3202  &$ -.124 $&  92.3916  &$ -.324 $&    94.2013  &$ -.475 $&   94.2725  &$ -.234 $&  94.3438  &$ -.260 $&  96.1378  &$ -.461 $\\
   92.2499  &$ -.460 $&     92.3212  &$ -.134 $&  92.3926  &$ -.327 $&    94.2023  &$ -.471 $&   94.2735  &$ -.218 $&  94.3447  &$ -.277 $&                    \\
   92.2509  &$ -.481 $&     92.3222  &$ -.121 $&  92.3936  &$ -.342 $&    94.2032  &$ -.483 $&   94.2745  &$ -.206 $&  94.3457  &$ -.280 $&                    \\
   92.2518  &$ -.478 $&     92.3232  &$ -.135 $&  92.3946  &$ -.346 $&    94.2042  &$ -.478 $&   94.2755  &$ -.208 $&  94.3467  &$ -.281 $&                    \\
   92.2528  &$ -.474 $&     92.3242  &$ -.110 $&  92.3956  &$ -.367 $&    94.2052  &$ -.462 $&   94.2765  &$ -.198 $&  94.3477  &$ -.274 $&                    \\
   92.2538  &$ -.471 $&     92.3252  &$ -.121 $&  92.3965  &$ -.368 $&    94.2062  &$ -.475 $&   94.2775  &$ -.199 $&  94.3487  &$ -.294 $&                    \\
\hline
\end{tabular}\end{tiny}
\end{table*}

\begin{table*}
\begin{tiny}
\caption{The original photometric data (HJD-2454400) of XY LMi in I
band.}
\begin{tabular}{llllllllllllll}
\hline
 HJD&$\Delta{m}$&HJD&$\Delta{m}$&HJD&$\Delta{m}$&HJD&$\Delta{m}$&HJD&$\Delta{m}$&HJD&$\Delta{m}$&HJD&$\Delta{m}$\\
 \hline
   92.1832  &$ -.416  $&    92.2550 &$  -.477  $& 92.3264  &$ -.153 $& 92.3978  &$ -.404 $& 94.2035  &$ -.483 $& 94.2747 &$  -.230 $&  94.3459   &$-.285$\\
   92.1847  &$ -.435  $&    92.2560 &$  -.485  $& 92.3274  &$ -.131 $& 92.3987  &$ -.370 $& 94.2044  &$ -.484 $& 94.2757 &$  -.224 $&  94.3469   &$-.298$\\
   92.1856  &$ -.416  $&    92.2570 &$  -.484  $& 92.3284  &$ -.144 $& 92.3997  &$ -.383 $& 94.2054  &$ -.483 $& 94.2767 &$  -.222 $&  94.3479   &$-.303$\\
   92.1866  &$ -.421  $&    92.2580 &$  -.481  $& 92.3294  &$ -.149 $& 92.4007  &$ -.466 $& 94.2064  &$ -.479 $& 94.2777 &$  -.211 $&  94.3489   &$-.309$\\
   92.1876  &$ -.434  $&    92.2590 &$  -.475  $& 92.3304  &$ -.139 $& 92.4017  &$ -.380 $& 94.2074  &$ -.492 $& 94.2787 &$  -.217 $&  94.3499   &$-.309$\\
   92.1886  &$ -.439  $&    92.2600 &$  -.478  $& 92.3313  &$ -.152 $& 92.4027  &$ -.486 $& 94.2084  &$ -.489 $& 94.2797 &$  -.208 $&  94.3509   &$-.329$\\
   92.1896  &$ -.429  $&    92.2610 &$  -.470  $& 92.3323  &$ -.146 $& 92.4037  &$ -.419 $& 94.2094  &$ -.485 $& 94.2806 &$  -.190 $&  94.3519   &$-.337$\\
   92.1906  &$ -.444  $&    92.2620 &$  -.468  $& 92.3333  &$ -.142 $& 92.4047  &$ -.491 $& 94.2104  &$ -.480 $& 94.2816 &$  -.181 $&  94.3529   &$-.324$\\
   92.1916  &$ -.435  $&    92.2630 &$  -.464  $& 92.3343  &$ -.139 $& 92.4057  &$ -.550 $& 94.2114  &$ -.482 $& 94.2826 &$  -.189 $&  94.3539   &$-.344$\\
   92.1926  &$ -.442  $&    92.2640 &$  -.463  $& 92.3353  &$ -.149 $& 92.4067  &$ -.417 $& 94.2124  &$ -.481 $& 94.2836 &$  -.182 $&  94.3548   &$-.339$\\
   92.1936  &$ -.452  $&    92.2649 &$  -.454  $& 92.3363  &$ -.131 $& 92.4077  &$ -.423 $& 94.2134  &$ -.485 $& 94.2846 &$  -.162 $&  94.3558   &$-.349$\\
   92.1946  &$ -.448  $&    92.2659 &$  -.455  $& 92.3373  &$ -.135 $& 92.4087  &$ -.425 $& 94.2143  &$ -.478 $& 94.2856 &$  -.171 $&  94.3568   &$-.351$\\
   92.1956  &$ -.454  $&    92.2669 &$  -.450  $& 92.3383  &$ -.146 $& 92.4096  &$ -.425 $& 94.2153  &$ -.467 $& 94.2866 &$  -.175 $&  94.3578   &$-.362$\\
   92.1966  &$ -.459  $&    92.2679 &$  -.458  $& 92.3393  &$ -.139 $& 92.4106  &$ -.447 $& 94.2163  &$ -.477 $& 94.2876 &$  -.161 $&  94.3588   &$-.355$\\
   92.1975  &$ -.446  $&    92.2689 &$  -.448  $& 92.3403  &$ -.133 $& 92.4116  &$ -.452 $& 94.2173  &$ -.467 $& 94.2886 &$  -.169 $&  94.3598   &$-.358$\\
   92.1985  &$ -.456  $&    92.2699 &$  -.438  $& 92.3413  &$ -.147 $& 92.4126  &$ -.447 $& 94.2183  &$ -.474 $& 94.2895 &$  -.168 $&  94.3608   &$-.386$\\
   92.1995  &$ -.462  $&    92.2709 &$  -.440  $& 92.3422  &$ -.146 $& 92.4136  &$ -.444 $& 94.2193  &$ -.455 $& 94.2905 &$  -.172 $&  94.3618   &$-.367$\\
   92.2005  &$ -.471  $&    92.2719 &$  -.435  $& 92.3432  &$ -.140 $& 92.4146  &$ -.455 $& 94.2203  &$ -.455 $& 94.2915 &$  -.162 $&  94.3628   &$-.380$\\
   92.2015  &$ -.463  $&    92.2729 &$  -.433  $& 92.3442  &$ -.159 $& 92.4156  &$ -.460 $& 94.2213  &$ -.462 $& 94.2925 &$  -.174 $&  94.3638   &$-.394$\\
   92.2025  &$ -.480  $&    92.2739 &$  -.428  $& 92.3452  &$ -.140 $& 92.4166  &$ -.432 $& 94.2223  &$ -.471 $& 94.2935 &$  -.165 $&  94.3647   &$-.379$\\
   92.2035  &$ -.488  $&    92.2749 &$  -.415  $& 92.3462  &$ -.154 $& 92.4176  &$ -.442 $& 94.2233  &$ -.470 $& 94.2945 &$  -.158 $&  94.3657   &$-.389$\\
   92.2045  &$ -.477  $&    92.2758 &$  -.421  $& 92.3472  &$ -.138 $& 92.4186  &$ -.529 $& 94.2242  &$ -.447 $& 94.2955 &$  -.175 $&  94.3667   &$-.395$\\
   92.2055  &$ -.461  $&    92.2768 &$  -.416  $& 92.3482  &$ -.140 $& 92.4196  &$ -.454 $& 94.2252  &$ -.449 $& 94.2965 &$  -.173 $&  94.3677   &$-.402$\\
   92.2065  &$ -.486  $&    92.2778 &$  -.416  $& 92.3492  &$ -.138 $& 92.4205  &$ -.474 $& 94.2262  &$ -.460 $& 94.2975 &$  -.167 $&  94.3687   &$-.400$\\
   92.2075  &$ -.482  $&    92.2788 &$  -.408  $& 92.3502  &$ -.135 $& 92.4215  &$ -.472 $& 94.2272  &$ -.455 $& 94.2985 &$  -.158 $&  94.3697   &$-.412$\\
   92.2085  &$ -.487  $&    92.2798 &$  -.403  $& 92.3512  &$ -.144 $& 92.4225  &$ -.486 $& 94.2282  &$ -.440 $& 94.2994 &$  -.154 $&  94.3707   &$-.409$\\
   92.2094  &$ -.489  $&    92.2808 &$  -.393  $& 92.3522  &$ -.135 $& 92.4235  &$ -.475 $& 94.2292  &$ -.434 $& 94.3004 &$  -.160 $&  94.3717   &$-.394$\\
   92.2104  &$ -.488  $&    92.2818 &$  -.408  $& 92.3531  &$ -.149 $& 92.4245  &$ -.568 $& 94.2302  &$ -.445 $& 94.3014 &$  -.165 $&  94.3727   &$-.400$\\
   92.2114  &$ -.506  $&    92.2828 &$  -.392  $& 92.3541  &$ -.143 $& 92.4255  &$ -.498 $& 94.2312  &$ -.402 $& 94.3024 &$  -.158 $&  94.3737   &$-.414$\\
   92.2124  &$ -.492  $&    92.2838 &$  -.368  $& 92.3551  &$ -.134 $& 92.4265  &$ -.455 $& 94.2322  &$ -.459 $& 94.3034 &$  -.164 $&  94.3747   &$-.410$\\
   92.2134  &$ -.492  $&    92.2848 &$  -.378  $& 92.3561  &$ -.141 $& 92.4275  &$ -.478 $& 94.2331  &$ -.417 $& 94.3044 &$  -.157 $&  94.3756   &$-.400$\\
   92.2144  &$ -.491  $&    92.2858 &$  -.373  $& 92.3571  &$ -.150 $& 92.4285  &$ -.516 $& 94.2341  &$ -.415 $& 94.3054 &$  -.159 $&  94.3766   &$-.396$\\
   92.2154  &$ -.503  $&    92.2867 &$  -.371  $& 92.3581  &$ -.142 $& 92.4295  &$ -.513 $& 94.2351  &$ -.434 $& 94.3064 &$  -.166 $&  94.3776   &$-.457$\\
   92.2164  &$ -.491  $&    92.2877 &$  -.366  $& 92.3591  &$ -.144 $& 92.4305  &$ -.474 $& 94.2361  &$ -.406 $& 94.3074 &$  -.163 $&  94.3786   &$-.419$\\
   92.2174  &$ -.501  $&    92.2887 &$  -.354  $& 92.3601  &$ -.142 $& 92.4314  &$ -.467 $& 94.2371  &$ -.426 $& 94.3084 &$  -.151 $&  94.3796   &$-.464$\\
   92.2184  &$ -.504  $&    92.2897 &$  -.350  $& 92.3611  &$ -.162 $& 92.4324  &$ -.413 $& 94.2381  &$ -.404 $& 94.3093 &$  -.155 $&  94.3806   &$-.429$\\
   92.2194  &$ -.515  $&    92.2907 &$  -.354  $& 92.3621  &$ -.174 $& 92.4334  &$ -.468 $& 94.2391  &$ -.412 $& 94.3103 &$  -.174 $&  94.3816   &$-.481$\\
   92.2203  &$ -.498  $&    92.2917 &$  -.343  $& 92.3631  &$ -.162 $& 92.4344  &$ -.508 $& 94.2401  &$ -.423 $& 94.3113 &$  -.168 $&  96.1053   &$-.351$\\
   92.2213  &$ -.506  $&    92.2927 &$  -.333  $& 92.3641  &$ -.177 $& 92.4354  &$ -.573 $& 94.2411  &$ -.407 $& 94.3123 &$  -.176 $&  96.1063   &$-.360$\\
   92.2223  &$ -.507  $&    92.2937 &$  -.326  $& 92.3650  &$ -.175 $& 92.4364  &$ -.497 $& 94.2421  &$ -.412 $& 94.3133 &$  -.160 $&  96.1073   &$-.369$\\
   92.2233  &$ -.513  $&    92.2947 &$  -.331  $& 92.3660  &$ -.186 $& 92.4374  &$ -.541 $& 94.2431  &$ -.404 $& 94.3143 &$  -.171 $&  96.1083   &$-.382$\\
   92.2243  &$ -.508  $&    92.2957 &$  -.319  $& 92.3670  &$ -.195 $& 92.4384  &$ -.499 $& 94.2440  &$ -.391 $& 94.3153 &$  -.149 $&  96.1093   &$-.370$\\
   92.2253  &$ -.505  $&    92.2967 &$  -.313  $& 92.3680  &$ -.202 $& 92.4394  &$ -.556 $& 94.2450  &$ -.381 $& 94.3163 &$  -.164 $&  96.1103   &$-.383$\\
   92.2263  &$ -.508  $&    92.2977 &$  -.310  $& 92.3690  &$ -.213 $& 92.4404  &$ -.480 $& 94.2460  &$ -.389 $& 94.3173 &$  -.166 $&  96.1113   &$-.379$\\
   92.2273  &$ -.506  $&    92.2986 &$  -.307  $& 92.3700  &$ -.207 $& 92.4414  &$ -.272 $& 94.2470  &$ -.382 $& 94.3183 &$  -.175 $&  96.1123   &$-.376$\\
   92.2283  &$ -.520  $&    92.2996 &$  -.295  $& 92.3710  &$ -.219 $& 92.4424  &$ -.556 $& 94.2480  &$ -.381 $& 94.3192 &$  -.163 $&  96.1133   &$-.392$\\
   92.2293  &$ -.518  $&    92.3006 &$  -.291  $& 92.3720  &$ -.237 $& 94.1777  &$ -.468 $& 94.2490  &$ -.394 $& 94.3202 &$  -.173 $&  96.1143   &$-.390$\\
   92.2303  &$ -.511  $&    92.3016 &$  -.291  $& 92.3730  &$ -.239 $& 94.1787  &$ -.468 $& 94.2500  &$ -.376 $& 94.3212 &$  -.160 $&  96.1152   &$-.404$\\
   92.2313  &$ -.515  $&    92.3026 &$  -.276  $& 92.3740  &$ -.239 $& 94.1797  &$ -.478 $& 94.2510  &$ -.363 $& 94.3222 &$  -.169 $&  96.1162   &$-.405$\\
   92.2322  &$ -.510  $&    92.3036 &$  -.270  $& 92.3750  &$ -.240 $& 94.1807  &$ -.480 $& 94.2520  &$ -.354 $& 94.3232 &$  -.157 $&  96.1172   &$-.388$\\
   92.2332  &$ -.507  $&    92.3046 &$  -.256  $& 92.3760  &$ -.259 $& 94.1817  &$ -.473 $& 94.2529  &$ -.369 $& 94.3242 &$  -.162 $&  96.1182   &$-.422$\\
   92.2342  &$ -.509  $&    92.3056 &$  -.263  $& 92.3769  &$ -.284 $& 94.1827  &$ -.492 $& 94.2539  &$ -.356 $& 94.3252 &$  -.164 $&  96.1192   &$-.416$\\
   92.2352  &$ -.508  $&    92.3066 &$  -.254  $& 92.3779  &$ -.279 $& 94.1837  &$ -.495 $& 94.2549  &$ -.360 $& 94.3262 &$  -.176 $&  96.1202   &$-.404$\\
   92.2362  &$ -.516  $&    92.3076 &$  -.242  $& 92.3789  &$ -.264 $& 94.1847  &$ -.461 $& 94.2559  &$ -.344 $& 94.3272 &$  -.175 $&  96.1212   &$-.423$\\
   92.2372  &$ -.519  $&    92.3086 &$  -.232  $& 92.3799  &$ -.295 $& 94.1857  &$ -.500 $& 94.2569  &$ -.337 $& 94.3281 &$  -.195 $&  96.1222   &$-.418$\\
   92.2382  &$ -.515  $&    92.3095 &$  -.212  $& 92.3809  &$ -.289 $& 94.1867  &$ -.494 $& 94.2579  &$ -.327 $& 94.3291 &$  -.181 $&  96.1232   &$-.428$\\
   92.2392  &$ -.515  $&    92.3105 &$  -.226  $& 92.3819  &$ -.321 $& 94.1876  &$ -.479 $& 94.2589  &$ -.323 $& 94.3301 &$  -.195 $&  96.1242   &$-.423$\\
   92.2402  &$ -.517  $&    92.3115 &$  -.208  $& 92.3829  &$ -.288 $& 94.1886  &$ -.483 $& 94.2599  &$ -.311 $& 94.3311 &$  -.210 $&  96.1251   &$-.421$\\
   92.2412  &$ -.505  $&    92.3125 &$  -.199  $& 92.3839  &$ -.282 $& 94.1896  &$ -.486 $& 94.2609  &$ -.330 $& 94.3321 &$  -.198 $&  96.1261   &$-.439$\\
   92.2422  &$ -.509  $&    92.3135 &$  -.266  $& 92.3849  &$ -.317 $& 94.1906  &$ -.477 $& 94.2619  &$ -.326 $& 94.3331 &$  -.193 $&  96.1271   &$-.440$\\
   92.2431  &$ -.506  $&    92.3145 &$  -.191  $& 92.3859  &$ -.402 $& 94.1916  &$ -.494 $& 94.2628  &$ -.306 $& 94.3341 &$  -.208 $&  96.1281   &$-.449$\\
   92.2441  &$ -.506  $&    92.3155 &$  -.192  $& 92.3869  &$ -.334 $& 94.1926  &$ -.490 $& 94.2638  &$ -.296 $& 94.3351 &$  -.191 $&  96.1291   &$-.447$\\
   92.2451  &$ -.500  $&    92.3165 &$  -.181  $& 92.3878  &$ -.336 $& 94.1936  &$ -.492 $& 94.2648  &$ -.299 $& 94.3361 &$  -.202 $&  96.1301   &$-.455$\\
   92.2461  &$ -.507  $&    92.3175 &$  -.181  $& 92.3888  &$ -.324 $& 94.1946  &$ -.500 $& 94.2658  &$ -.302 $& 94.3370 &$  -.207 $&  96.1311   &$-.464$\\
   92.2471  &$ -.506  $&    92.3185 &$  -.157  $& 92.3898  &$ -.355 $& 94.1956  &$ -.482 $& 94.2668  &$ -.275 $& 94.3380 &$  -.229 $&  96.1321   &$-.460$\\
   92.2481  &$ -.504  $&    92.3195 &$  -.173  $& 92.3908  &$ -.342 $& 94.1965  &$ -.482 $& 94.2678  &$ -.271 $& 94.3390 &$  -.239 $&  96.1331   &$-.479$\\
   92.2491  &$ -.510  $&    92.3205 &$  -.164  $& 92.3918  &$ -.337 $& 94.1975  &$ -.493 $& 94.2688  &$ -.278 $& 94.3400 &$  -.242 $&  96.1341   &$-.459$\\
   92.2501  &$ -.492  $&    92.3214 &$  -.148  $& 92.3928  &$ -.378 $& 94.1985  &$ -.493 $& 94.2698  &$ -.256 $& 94.3410 &$  -.250 $&  96.1351   &$-.456$\\
   92.2511  &$ -.491  $&    92.3224 &$  -.144  $& 92.3938  &$ -.502 $& 94.1995  &$ -.496 $& 94.2708  &$ -.253 $& 94.3420 &$  -.255 $&  96.1360   &$-.461$\\
   92.2521  &$ -.498  $&    92.3234 &$  -.156  $& 92.3948  &$ -.370 $& 94.2005  &$ -.480 $& 94.2717  &$ -.260 $& 94.3430 &$  -.268 $&  96.1370   &$-.466$\\
   92.2531  &$ -.488  $&    92.3244 &$  -.132  $& 92.3958  &$ -.451 $& 94.2015  &$ -.485 $& 94.2727  &$ -.245 $& 94.3440 &$  -.255 $&  96.1380   &$-.465$\\
   92.2540  &$ -.502  $&    92.3254 &$  -.144  $& 92.3968  &$ -.364 $& 94.2025  &$ -.502 $& 94.2737  &$ -.242 $& 94.3450 &$  -.280 $&            &$     $\\
\hline
\end{tabular}\end{tiny}
\end{table*}

Complete light curves in B, V, R, and I bands were obtained and are
shown in Figure 1. In all, 584 data points in B, 585 in V, 575 in R,
and 565 in I were obtained. The corresponding observational data,
i.e., the original HJD date and magnitude difference between XY LMi
and the comparison star, are listed in Tables 2, 3, 4, and 5. As
shown in Fig. 1, the light curves in 4 colors are typical EW-type
where light variation is continuous and has a very small difference
between the depths of the two minima. These properties reveal
tidally distorted components and both of the components have similar
temperature. The amplitudes of the light variation are
$\approx0.401$\,mag in B band, $\approx0.383$\,mag in V band,
$\approx0.363$\,mag in R band, $\approx0.346$\,mag in I band. As in
the light curve obtained by B\&B, the eclipses are complete
indicating that the BVRI light curves are very useful to determine
reliable photometric parameters of the system. Therefore, to derive
photometric elements and to understand the evolutionary state of the
binary star, the present light curves were analyzed simultaneously
with the 2003 version of the W-D program (Wilson \& Devinney 1971;
Wilson 1979, 1990).

\begin{figure}
\begin{center}
\includegraphics[angle=0,scale=1.2]{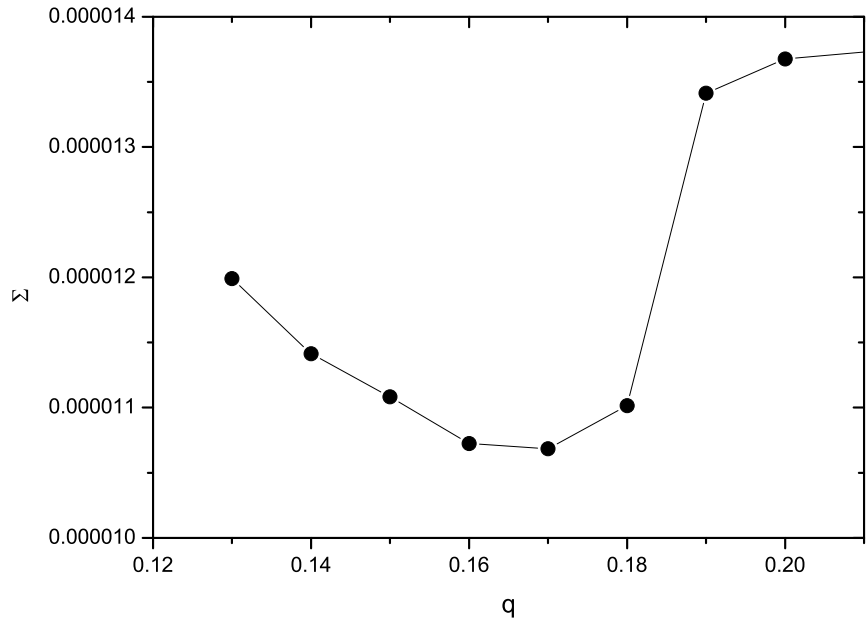}
\caption{The relation between $\Sigma$ and q for XY Leonis Minoris.}
\end{center}
\end{figure}

The color indexes of the binary star given by Gettel et al. (2006)
and Cutri et al. (2003) are: B-V=0.525; J-H=0.222; H-K=0.049;
V-K=1.399, which corresponds different values of temperature for the
primary component (Cox, 2000). Therefore, we choose a mean value,
i.e., the temperature for star 1 (star eclipsed at primary light
minimum) was fixed as $T_1=6144\,K$. During the analyzing, all of
original individual data points were used. We assumed convective
outer envelopes for both components, and the bolometric albedo
$A_{1}=A_{2}=0.5$ (Rucinski 1969) and the values of the
gravity-darkening coefficients $g_{1}=g_{2}=0.32$ (Lucy 1967) were
used. The limb-darkening coefficients for different filters were
taken from van Hamme's table (1993). The adjustable parameters were:
the orbital inclination i; the mean temperature of star 2, $T_{2}$;
the mass ratio $q$($q=M_2/M_1$); the band-pass luminosity of star 1,
$L_{1B}$ and $L_{1V}$; and the dimensionless potential of star 1
($\Omega_{1}=\Omega_{2}$, mode 3 for overcontact configuration).

Since no mass ratios of XY LMi are in the previous literature, a
q-search method was used to determine its mass ratio. The total and
shallow eclipsing minima in the light curves indicate that it is a
low mass ratio contact binary with mass ratio between 0.1 and 0.2
(e.g., Wilson, 1978; Terrell \& Wilson, 2005; Wilson, 2006).
Therefore, we focus on searching for photometric solutions with mass
ratio below 0.2, and solutions were carried out for a series of
values of the mass ratio (q=0.13, 0.14, 0.15, 0.16, 0.17, 0.18,
0.19, and 0.20). For each value of q, the calculation started at
mode 2 (detached mode) and we discovered that the solutions usually
converged to overcontact configuration. The relation between the
resulting sum $\Sigma$ of weighted square deviations and q is
plotted in Fig. 2. A minimum value was obtained at q=0.16.
Therefore, we chose the initial value of mass ration q as 0.16 and
made it an adjustable parameter. Then, we performed a differential
correction until it converged and final solutions were derived.

\begin{table}
\caption{Photometric solutions for XY Leonis Minoris.}
\begin{tabular}{lll}
\tableline
 Parameters & Without spots & With dark spots\\\tableline
$g_{1}=g_{2}$&0.32& 0.32 \\
$A_{1}=A_{2}$&0.5 & 0.5  \\
$x_{1B}=x_{2B}$ & 0.690 & 0.690 \\
$x_{1V}=x_{2V}$ & 0.565 & 0.565 \\
$x_{1R}=x_{2R}$ & 0.467 & 0.467 \\
$x_{1I}=x_{2I}$ & 0.380 & 0.380 \\
$T_{1}$ & 6144K & 6144K \\
q   & $0.1505(\pm0.0007)$ & $0.1480(\pm0.0008)$\\
$\Omega_{in}$ & 2.10309   & 2.09755\\
$\Omega_{out}$& 2.00627   & 2.00200\\
$T_{2}$ & $6089(\pm6)$K & $6093(\pm6)$K \\
$i$     & $81.50(\pm0.40)$ & $81.04(\pm0.35)$\\
$\Omega_{1}=\Omega_{2}$ & $2.0268(\pm0.0034)$ & $2.0268(\pm0.0034)$ \\
$L_{1}/(L_{1}+L_{2})$ (B) &$0.8394(\pm0.0002)$ & $0.8390(\pm0.0002)$\\
$L_{1}/(L_{1}+L_{2}$) (V) &$0.8379(\pm0.0002)$ & $0.8377(\pm0.0001)$\\
$L_{1}/(L_{1}+L_{2})$ (R) &$0.8371(\pm0.0001)$ & $0.8370(\pm0.0001)$\\
$L_{1}/(L_{1}+L_{2}$) (I) &$0.8364(\pm0.0001)$ & $0.8363(\pm0.0001)$\\
$r_{1}(pole)$&$0.5248(\pm0.0010)$ & $0.5275(\pm0.0010)$\\
$r_{1}(side)$&$0.5840(\pm0.0015)$ & $0.5880(\pm0.0015)$\\
$r_{1}(back)$&$0.6101(\pm0.0019)$ & $0.6145(\pm0.0019)$\\
$r_{2}(pole)$&$0.2350(\pm0.0020)$ & $0.2358(\pm0.0021)$\\
$r_{2}(side)$&$0.2476(\pm0.0025)$ & $0.2488(\pm0.0026)$\\
$r_{2}(back)$&$0.3104(\pm0.0083)$ & $0.3171(\pm0.0098)$\\
The degree of overcontact (f) & $66.6(\pm3.6\,\%)$ & $74.1(\pm3.6\,\%)$\\
\tableline
\end{tabular}
\end{table}

\begin{figure}
\begin{center}
\includegraphics[angle=0,scale=1.1]{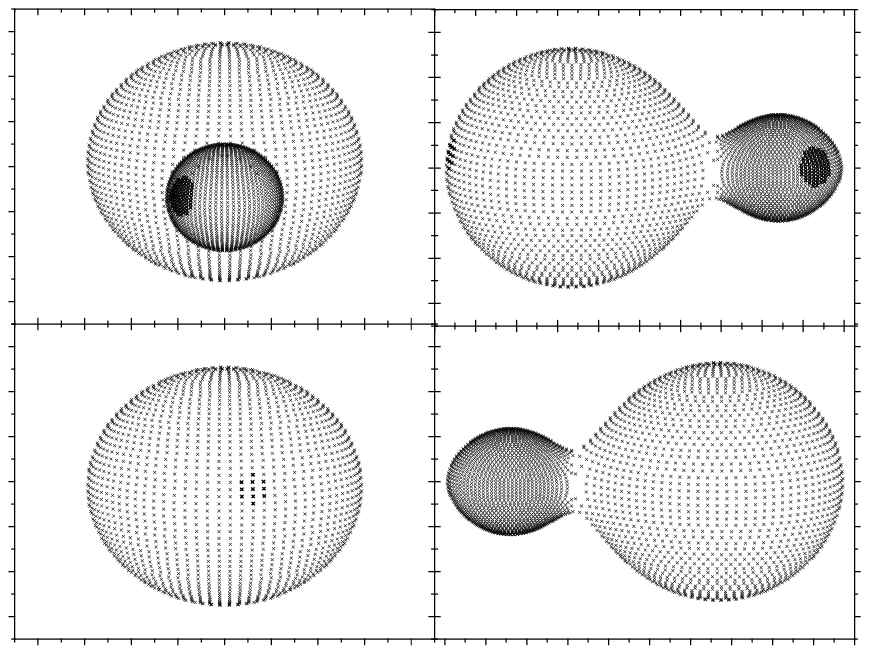}
\caption{Geometrical structure of XY Leonis Minoris with one dark
spot on the primary and one on the secondary at phases of 0.00,
0.25, 0.50, and 0.75.}
\end{center}
\end{figure}

The photometric solutions are listed in Table 6. However, we found
that the theoretical light curves do not fit the observations very
good between phases 0.65 and 0.80. Both components of the binary are
cool stars and shares a common convective envelope. The deep
convective envelope along with fast rotation can produce a strong
magnetic dynamo and solar-like magnetic activity including
photospheric dark spots. Therefore, the disagreement between the
theoretical light curve and the observation was explained as the
presence of dark spots on the common convective envelope. Two dark
spots were introduced , one on each star. In the 2003 version of the
W-D program, each spot was described by four parameters: spot center
longitude ($\theta$), spot center latitude ($\phi$), spot angular
radius ($r$) (all in units of radian), and the spot temperature
factor $T_f=T_d/T_0$, the ratio of local spot temperature to
unspotted temperature. The photometric solutions with two dark spots
are listed in Table 6, and the parameters of the two dark spots are
shown in Table 7. The corresponding theoretical light curves are
plotted in Fig. 1 that fit the observations better. The geometrical
structures of XY LMi at phases of 0.00, 0.25, 0.50, and 0.75, are
displayed in Fig. 3. Our photometric solutions suggest that XY LMi
is a deep-overcontact binary system with a high degree of
overcontact ($f=74.1\,\%$). The derived mass ratio is $q=0.148$.

Recent investigations have shown that nearly 100\% of very close
binaries ($P<10$\,days), especially W UMa-type binary stars (with
the shortest periods among main-sequence binaries of the same
spectral types), are members of triple systems (e.g. Tokovinin, et
al., 2006; Pribulla \& Rucinski, 2006; D¡¯Angelo, et al., 2006; Qian
et al., 2006). To check on possible third light, during the solution
we also included third light as one of the adjusted parameters.
However, the results suggest that third light is negligible and
negative indicating that, if there is a third body companion to XY
LMi, it should be an extremely faint object.

\begin{table}
\caption{Parameters of the dark spots on the photospheric surfaces
of the two components.}
\begin{tabular}{llll}\hline
Parameters  & Spot 1 (on the primary) & Spot 2 (on the
secondary)\\\hline
$\theta$(radian)  & 1.4378  & 1.4377 \\
$\phi$(radian)   & 3.3554  & 2.3555 \\
$r$(radian)       & 0.1247  & 0.3247 \\
$T_f(T_d/T_0)$    & 0.8333  & 0.7033 \\
\hline
\end{tabular}
\tablecomments{Details of explanations on the spot center longitude
($\theta$) and latitude ($\phi$) were given by Wilson \& Van Hamme
(2003).}
\end{table}

\section{Orbital period variation of XY Leonis Minoris}

\begin{table}
\caption{New determined CCD times of light minimum for XY Leonis
Minoris}
\begin{center}
\begin{tabular}{llllll}\hline\hline
J.D. (Hel.) (days) &Error (days) &Methods& Min. & Filter &
Telescopes\\\hline
 2452342.4222 & $\pm0.0022$ & CCD & I     & Non & The 21-cm  \\
 2452342.6357 & $\pm0.0005$ & CCD & II    & Non & The 21-cm  \\
 2452346.4648 & $\pm0.0006$ & CCD & II    & Non & The 21-cm  \\
 2452381.5203 & $\pm0.0006$ & CCD & II    & Non & The 21-cm  \\
 2452382.3936 & $\pm0.0009$ & CCD & II    & Non & The 21-cm  \\
 2452383.4869 & $\pm0.0004$ & CCD & I     & Non & The 21-cm  \\
 2452615.6901 & $\pm0.0009$ & CCD & II    & Non & The 21-cm  \\
 2452734.5230 & $\pm0.0010$ & CCD & II    & Non & The 21-cm  \\
 2453739.3720 & $\pm0.0015$ & CCD & II    & V   & The 60-cm  \\
 2453739.3638 & $\pm0.0012$ & CCD & II    & R   & The 60-cm  \\
 2453766.2260 & $\pm0.0010$ & CCD & I     & B   & The 60-cm  \\
 2453766.2277 & $\pm0.0011$ & CCD & I     & V   & The 60-cm  \\
 2453766.2319 & $\pm0.0005$ & CCD & I     & R   & The 60-cm  \\
 2454492.3408 & $\pm0.0002$ & CCD & I     & B   & The 85-cm  \\
 2454492.3407 & $\pm0.0002$ & CCD & I     & V   & The 85-cm  \\
 2454492.3408 & $\pm0.0002$ & CCD & I     & R   & The 85-cm  \\
 2454492.3409 & $\pm0.0003$ & CCD & I     & I   & The 85-cm  \\
 2454494.3054 & $\pm0.0003$ & CCD & II    & B   & The 85-cm  \\
 2454494.3057 & $\pm0.0002$ & CCD & II    & V   & The 85-cm  \\
 2454494.3062 & $\pm0.0003$ & CCD & II    & R   & The 85-cm  \\
 2454494.3061 & $\pm0.0002$ & CCD & II    & I   & The 85-cm  \\
 2454574.0379 & $\pm0.0014$ & CCD & I     & R   & The 1.0-m  \\
 2454809.3015 & $\pm0.0002$ & CCD & II    & R   & The 1.0-m  \\
 2454887.2866 & $\pm0.0002$ & CCD & I     & R   & The 1.0-m  \\
 2454937.0903 & $\pm0.0002$ & CCD & I     & R   & The 60-cm  \\
 2455296.2123 & $\pm0.0011$ & CCD & I     & R   & The 60-cm  \\
 2455311.0660 & $\pm0.0006$ & CCD & I     & R   & The 60-cm  \\
 \hline
\end{tabular}
\end{center}
\tablecomments{Explanations on the used telescopes are as following.
The 21-cm: the 21.2-cm telescope in Les Engarouines Observatory; The
60-cm: the 60-cm telescope in Yunnan Astronomical Observatory; The
85-cm: the 85-cm telescope in Xinglong station of National
Astronomical Observatories; The 1.0-m: the 1.0-m telescope in Yunnan
Astronomical Observatory.}
\end{table}

After the first linear ephemeris of XY LMi derived by B\&B, only
three eclipse times of the eclipsing binary was obtained by
Maciejewski \& Karska (2004) and by Hubscher et al. (2009). By using
the photometric data published by B\&B, 8 times of minimum light
were determined and are listed in Table 8. It has shown in previous
papers (e.g., Qian \& Yang (2004); Zhu et al. (2005); Qian et al.
(2005a, b, 2006, 2007, 2008)) that the orbital periods of all of the
investigated high fill-out, extreme mass ratio overcontact binaries
are variable. To check the orbital period of XY LMi is variable or
not, three small telescopes, e.g., the 60-cm and the 1.0-m
telescopes in Yunnan Astronomical Observatory and the 80-cm
telescope in Xinglong station of National Astronomical
Observatories, were used to monitor it for determining eclipse
times. With our data, 19 times of light minimum were determined by
using the parabolic fitting method and are shown in Table 8.

The timing residuals (O-C values) of all available eclipse times
were calculated with the linear ephemeris determined by B\&B (see in
Table 9). The corresponding timing-residual diagram is plotted
against epoch number E in Fig. 4. It is shown in the upper panel of
Fig. 4 that the B\&B period needed refinement (as the plotted
relation is not close to horizontal), and the period of XY LMi is
variable. By assuming a continuously long-term period change, a
least-squares solution leads to the following equation,
\begin{eqnarray}
Min.I &=&2452366.88397(\pm0.00002)+0.43688773(\pm0.00000002)\times{E}\nonumber\\
         & &-1.00(\pm0.03)\times{10^{-10}}\times{E^{2}}.
\end{eqnarray}
The quadratic term in this equation reveals a period decrease at a
rate of $dP/dt=-1.67\times{10^{-7}}$\,days/year which is the same
order as those determined in other overcontact binaries (e.g., Qian
2001, 2003a, b). After the long-term period change was removed, the
residuals respect to the quadratic ephemeris are shown in the lowest
panel where no changes can be traced indicating that the quadratic
ephemeris can fit timing-residual curve very well.

The continuous period decrease may be caused by angular momentum
loss (AML) or by a combination of the mass transfer from the primary
to the secondary and angular momentum loss via magnetic stellar
wind. As the period is decreasing, the inner and outer critical
Roche lobes will be shrinking and thus will cause $f$ increasing.
Finally, as in the cases of GR Vir (Qian \& Yang, 2004), FG Hya
(Qian \& Yang, 2005), IK Per (Zhu et al., 2005), CU Tauri, and TV
Muscae (Qian et al. 2005a), it will evolve into a single
rapid-rotation star before the fluid surface reaching the outer
critical Roche Lobe (Rasio \& Shapiro 1995). However, the continuous
period decrease may be only a part of a long-period cyclic variation
or a combination of a cyclic change and a long-term variation as
have been observed in other W UMa-type binaries (e.g., V417 Aql,
Qian 2003b). Moreover, as shown in Fig. 4, the time interval of
eclipse times is only eight years. To check the period change of XY
LMi presented here, more times of light minimum are required in the
future.

\begin{figure}
\begin{center}
\includegraphics[angle=0,scale=1.1]{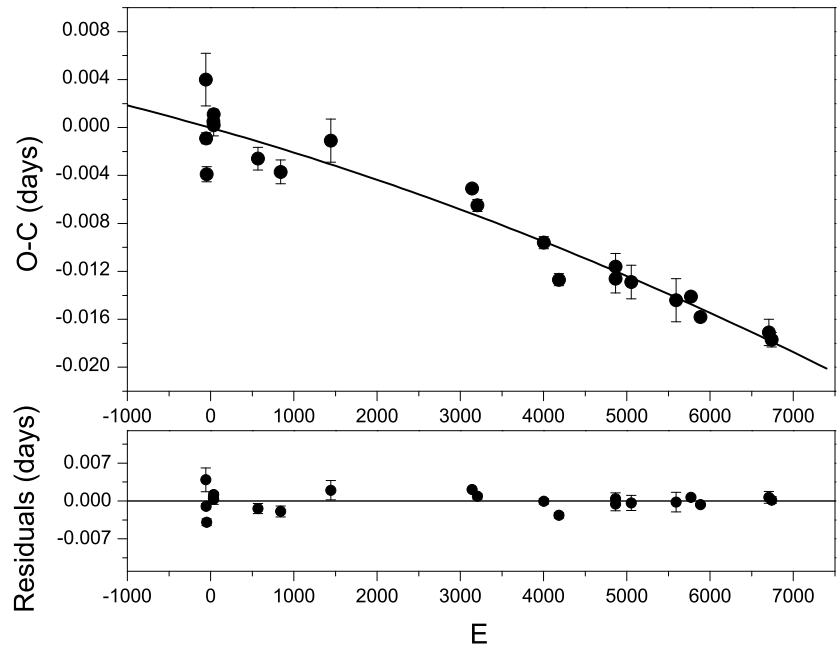}
\caption{$O-C$ diagram of XY Leonis Minoris (upper panel) from the
linear ephemeris of Eq. (1). The solid line in the upper panel
refers to a continuous period decrease. The residuals respect to the
quadratic ephemeris are shown in the lowest panel where no changes
can be traced.}
\end{center}
\end{figure}

%\begin{table}
%\begin{footnotesize}
\begin{deluxetable}{llllll}
\tablewidth{0pc} \tabletypesize{\footnotesize} \tablecaption{O-C
values of the eclipse times of XY Leonis Minoris.}
%\begin{tabular}{llllllll}
%\hline
\tablehead{ \colhead{JD.Hel.}  & \colhead{Min.}  & \colhead{E}  &
\colhead{$O-C$} & \colhead{Residuals}& \colhead{Ref.}}
%\hline
\startdata
2400000+   &    &               &      &   &\\
 2452342.4222 & I  & -56     & +0.0040 & +0.0039 & (1)\\
 2452342.6357 & II & -55.5   & -0.0009 & -0.0010 & (1)\\
 2452346.4648 & II & -46.5   & -0.0039 & -0.0040 & (1)\\
 2452381.5203 & II & 33.5    & +0.0005 & +0.0006 & (1)\\
 2452382.3936 & II & 35.5    & +0.0002 & +0.0003 & (1)\\
 2452383.4869 & I  & 38      & +0.0011 & +0.0012 & (1)\\
 2452615.6901 & II & 569.5   & -0.0026 & -0.0014 & (1)\\
 2452734.5230 & II & 841.5   & -0.0037 & -0.0019 & (1)\\
 2452997.9701 & II & 1444.5  & -0.0011 & +0.0020 & (2)\\
 2453739.3679 & II & 3141.5  & -0.0051 & +0.0021 & (1)\\
 2453766.2352 & I  & 3203    & -0.0065 & +0.0009 & (1)\\
 2454115.5254 & II & 4002.5  & -0.0096 & -0.0006 & (3)\\
 2454195.4731 & II & 4185.5  & -0.0127 & -0.0027 & (3)\\
 2454492.3408 & I  & 4865    & -0.0116 & +0.0004 & (1)\\
 2454494.3058 & II & 4869.5  & -0.0126 & -0.0006 & (1)\\
 2454574.0379 & I  & 5052    & -0.0129 & -0.0003 & (1)\\
 2454809.3015 & II & 5590.5  & -0.0144 & -0.0002 & (1)\\
 2454887.2866 & I  & 5769    & -0.0141 & +0.0007 & (1)\\
 2454937.0903 & I  & 5883    & -0.0158 & -0.0007 & (1)\\
 2455296.2123 & I  & 6705    & -0.0171 & +0.0007 & (1)\\
 2455311.0660 & I  & 6739    & -0.0177 & +0.0002 & (1)\\
\enddata
%\hline
%\end{tabular}
%\end{footnotesize}\\
\tablecomments{References in Table 9:\\
(1) The present authors; (2) Maciejewski \& Karska (2004); (3)
Hubscher et al. (2009).}
\end{deluxetable}

\acknowledgments{This work is partly supported by Chinese Natural
Science Foundation (No.10973037, No.10903026, and No.10878012), the
National Key Fundamental Research Project through grant
2007CB815406, the Yunnan Natural Science Foundation (No. 2008CD157),
and by West Light Foundation of the Chinese Academy of Sciences. New
CCD photometric observations of the system were obtained with the
1.0-m and 60-cm telescopes at Yunnan Observatory and the 85-cm
telescope at Xinglong station of NAO.}

\end{document}